\documentclass[12pt]{article}
\usepackage{amsmath}
\usepackage{graphicx,psfrag,epsf}
\usepackage{enumerate}
\usepackage{natbib}
\usepackage{url}
\usepackage{amsmath,amssymb,amsfonts,amsthm,mathtools}
\usepackage{bm,bbm}
\usepackage[colorlinks=true, citecolor=blue, linkcolor=blue, urlcolor=blue]{hyperref}
\usepackage{color}
\usepackage{subfigure}
\usepackage{algorithm,algpseudocode}
\usepackage{scalefnt}
\usepackage{rotating}
\usepackage{adjustbox}

\newtheorem{rem}{{Remark}}

\newtheorem{prop}{{Proposition}}

\begin{document}

\def\spacingset#1{\renewcommand{\baselinestretch}%
{#1}\small\normalsize} \spacingset{1}

\title{\bf A multivariate Birnbaum-Saunders autoregressive moving average model with application to air pollution concentration data}

\author{
  \large
  {{Helton Saulo}$^{1,2}$\footnote{Corresponding author: heltonsaulo@gmail.com.}}\\[-0.05cm]
  {\small $^{1}$Department of Statistics, University of Brasilia, Brasilia, Brazil}\\[-0.05cm]
  {\small $^{2}$Department of Economics, Federal University of Pelotas, Pelotas, Brazil}\\[-0.05cm]
}

\date{}

\maketitle

\bigskip
\vspace{-0.5cm}
\begin{abstract}
Fine particulate matter (PM$_{2.5}$) concentration data are positive,
right-skewed series that arise naturally in environmental monitoring and are
well described by the Birnbaum-Saunders (BS) distribution. In this paper, we propose a multivariate BS autoregressive
moving average (MBSARMA) model with exogenous terms for the joint analysis of
correlated positive asymmetric time series. The proposed model combines the
multivariate log-linear BS framework with dynamic autoregressive moving average
components on the conditional location parameter of each response. We
estimate the model parameters by means of the Expectation-Maximisation
(EM) algorithm. The performance of the proposed conditional likelihood estimators is
evaluated by means of a Monte Carlo simulation study under several correlation levels and sample sizes. An application to weekly
PM$_{2.5}$ pollution concentration data recorded at three monitoring stations
in Santiago, Chile, obtained from the National Air Quality Information System
of Chile (SINCA), is presented. The results show the good performance of the
proposed methodology.
\end{abstract}

\noindent%

\noindent%
\noindent%

\section{Introduction}\label{sec:01}

The Birnbaum-Saunders (BS) distribution was introduced by
\cite{birnbaumsaunders:69} as a life distribution to describe the failure time
of materials subject to cyclical stress. The BS distribution is unimodal,
positively skewed, and defined on the positive real line, with two parameters
controlling its shape and scale. It has been widely applied in diverse fields such as engineering, environmental, and
medical sciences \citep{johnson:94,leiva:16}. In particular,
\cite{riecknedelman:91} introduced the log-linear
version of the BS distribution, known as the log-BS distribution, which is
useful for formulating regression and time series models. The log-BS
distribution is characterized by different shapes ranging from unimodal to bimodal depending on
its parameters, making it a
practical alternative to the normal distribution.

Autoregressive moving average (ARMA) models based on the BS distribution have
been proposed for the analysis of positive, asymmetric time series. In the
univariate setting, \cite{leiva:21} introduced the BS autoregressive moving average
(BISARMA) model, which incorporates autoregressive and moving average components
together with exogenous regressors within the log-BS framework, and established
its stationarity, invertibility, and estimation properties. In the context of duration models,
\cite{b:10} introduced the BS autoregressive conditional
duration (ACD) model, whereas \cite{lslm:14} proposed a broader family of ACD
models incorporating the BS distribution alongside other positive laws.
\cite{saulolla:19} applied BS-ACD models to high-frequency financial data,
and \cite{sauloetal:23} proposed quantile skew-BS ACD models for duration analysis.

The multivariate extension of the BS distribution has received significant
attention in the literature. \cite{marchant:15} proposed the multivariate
log-linear model for BS distributions, which provides a general framework for
jointly modeling correlated positive asymmetric responses using the theory of
elliptically contoured distributions and the expectation-maximization (EM) algorithm
for parameter estimation. Building on that framework, \cite{marchant:17}
developed robust multivariate control charts based on BS distributions,
illustrating the methodology with real-world air quality data from Chile.
These contributions demonstrate the practical relevance of the multivariate BS
family in diverse applied settings.

Environmental monitoring of air quality constitutes an important applied
context in which the BS distribution proves particularly useful. Fine
particulate matter (PM$_{2.5}$) concentration data are defined on the
positive real line and typically exhibit right-skewed, heavy-tailed behavior,
especially at the higher concentration levels that are characteristic of
pollution episodes. The suitability of the BS distribution for environmental
data was established by \cite{leiva:08}, who proposed generalized BS
distributions and illustrated their use with air pollutant concentration
series, and further reinforced by \cite{leiva:15env}, who developed a BS
attribute control chart for environmental assessment. In particular,
\cite{puentes:21} showed that bivariate BS log-linear models provide an
effective tool for jointly predicting PM$_{2.5}$ and PM$_{10}$ levels during
critical pollution episodes in Santiago, Chile, whereas \cite{ibacache:26}
proposed a partially linear varying coefficient model based on the BS
distribution to analyze PM$_{2.5}$ concentrations, explicitly confirming that
the positive skewness and non-negative support of PM$_{2.5}$ series are
consistent with the BS distributional framework.

In general, the recording of PM$_{2.5}$ concentrations are carried out simultaneously at multiple
monitoring stations within the same urban area and may exhibit strong correlation,
a feature that cannot be adequately captured by univariate
models and that motivates the multivariate framework proposed here. However,
to the best of our knowledge, no work has combined the multivariate BS
framework with dynamic ARMA components to produce a model capable of jointly
handling correlated multivariate PM$_{2.5}$ concentration time series. In order to fill this gap, the main objective of this paper is to propose and investigate the MBSARMA model. The secondary objectives are: (a) to estimate the model parameters by means of the EM algorithm; (b) to carry out a Monte Carlo simulation study to evaluate
the performance of the corresponding conditional maximum likelihood (ML) estimators; and
(c) to illustrate the proposed methodology with an application to real-world
multivariate air pollution concentration data.

The paper is organized as follows. In Section~\ref{sec:02}, we describe the
BS and multivariate log-BS distributions. Section~\ref{sec:03} introduces the
MBSARMA model formulation, statistical properties, parameter estimation,
prediction, and residual analysis. Section~\ref{sec:04} presents the Monte
Carlo simulation study. Section~\ref{sec:05} presents an application to weekly PM$_{2.5}$ air
pollution concentration data. Finally, Section~\ref{sec:06} presents some
concluding remarks.

\section{Preliminaries}\label{sec:02}
This section provides a brief overview of the univariate BS and log-BS distributions, along with the multivariate log-BS distribution, which serve as the foundation for the proposed MBSARMA model.
\subsection{The Birnbaum-Saunders distribution}\label{sec:021}

A random variable $T$ follows a BS distribution with shape parameter
$\alpha > 0$ and scale parameter $\beta > 0$, denoted $T \sim \text{BS}(\alpha,
\beta)$, if it can be represented as
\begin{equation}\label{eq:bsrep}
  T = \beta\!\left(\frac{\alpha Z}{2} + \sqrt{\left(\frac{\alpha Z}{2}\right)^2
  + 1}\right)^{\!2},
\end{equation}
where $Z \sim N(0,1)$; see \cite{birnbaumsaunders:69}. From
equation~(\ref{eq:bsrep}), the cumulative distribution function (CDF) and the
probability density function (PDF) of $T$ are expressed, respectively, as
\begin{equation}\label{eq:bscdf}
  F_T(t; \alpha, \beta)
  = \Phi\!\left[\frac{1}{\alpha}\!\left(\sqrt{\frac{t}{\beta}}
    - \sqrt{\frac{\beta}{t}}\right)\right], \quad t > 0,
\end{equation}
and
\begin{equation}\label{eq:bspdf}
  f_T(t; \alpha, \beta)
  = \frac{1}{2\alpha\sqrt{2\pi}\,\beta}\!\left[\left(\frac{\beta}{t}
    \right)^{1/2} + \left(\frac{\beta}{t}\right)^{3/2}\right]
    \exp\!\left[-\frac{1}{2\alpha^2}\!\left(\frac{t}{\beta}
    + \frac{\beta}{t} - 2\right)\right], \quad t > 0,
\end{equation}
where $\Phi(\cdot)$ is the standard normal CDF. It is relevant to note that
the scale parameter $\beta$ is the median of $T \sim \text{BS}(\alpha, \beta)$,
and that $kT \sim \text{BS}(\alpha, k\beta)$ for any constant $k > 0$, so the
BS distribution is closed under proportional changes of scale.

The log-BS distribution arises naturally as the logarithm of a BS random
variable. If $T \sim \text{BS}(\alpha, \beta)$, then
$Y = \log(T) \sim \log\text{-BS}(\alpha, \mu)$ with $\mu = \log(\beta)
\in \mathbb{R}$; see \cite{riecknedelman:91}. The CDF and PDF of $Y$ are
given, respectively, as
\begin{equation}\label{eq:logbscdf}
  F_Y(y; \alpha, \mu)
  = \Phi\!\left[\frac{2}{\alpha}\sinh\!\left(\frac{y - \mu}{2}\right)\right],
  \quad y \in \mathbb{R},
\end{equation}
and
\begin{equation}\label{eq:logbspdf}
  f_Y(y; \alpha, \mu)
  = \frac{1}{\alpha\sqrt{2\pi}}
    \exp\!\left[-\frac{2}{\alpha^2}\sinh^2\!\left(\frac{y-\mu}{2}\right)\right]
    \cosh\!\left(\frac{y-\mu}{2}\right), \quad y \in \mathbb{R},
\end{equation}
where $\alpha > 0$ is the shape parameter and $\mu \in \mathbb{R}$ is the
location parameter. An important property follows directly from (\ref{eq:logbscdf}):
$Y \sim \log\text{-BS}(\alpha, \mu)$ if and only if
\begin{equation}\label{eq:zscore}
  Z = \frac{2}{\alpha}\sinh\!\left(\frac{Y - \mu}{2}\right) \sim N(0,1).
\end{equation}
This characterization is fundamental for generating random variates, for
formulating time series models, and for constructing residuals; see
\cite{leiva:21}.

\subsection{The multivariate log-Birnbaum-Saunders distribution}\label{sec:022}

Let $\mathbf{Y} = (Y_1, \ldots, Y_d)^\top \in \mathbb{R}^d$ be a random
vector. In order to define the multivariate log-BS distribution, we write
the component-wise transformations
\begin{equation}\label{eq:adef}
  a_j = \frac{2}{\alpha}\sinh\!\left(\frac{y_j - \mu_j}{2}\right),
  \quad
  c_j = \frac{2}{\alpha}\cosh\!\left(\frac{y_j - \mu_j}{2}\right),
  \quad j = 1, \ldots, d,
\end{equation}
where $\alpha > 0$ and $\boldsymbol{\mu} = (\mu_1, \ldots, \mu_d)^\top \in
\mathbb{R}^d$. The $d$-variate log-BS distribution with common shape parameter
$\alpha$, location vector $\boldsymbol{\mu}$, and correlation matrix
$\boldsymbol{\Psi} \in \mathbb{R}^{d \times d}$, denoted $\mathbf{Y} \sim
\log\text{-BS}_d(\alpha, \boldsymbol{\mu}, \boldsymbol{\Psi})$, is defined by
the requirement that $\mathbf{a} = (a_1, \ldots, a_d)^\top \sim
N_d(\mathbf{0}, \boldsymbol{\Psi})$; see \cite{marchant:15}. The joint PDF of
$\mathbf{Y}$ takes the form
\begin{equation}\label{eq:mlbspdf}
  f_{\mathbf{Y}}(\mathbf{y};\alpha,\boldsymbol{\mu},\boldsymbol{\Psi})
  = \frac{1}{(2\pi)^{d/2} \cdot 2^d \cdot |\boldsymbol{\Psi}|^{1/2}}
    \exp\!\left(-\frac{1}{2}\,\mathbf{a}^\top\boldsymbol{\Psi}^{-1}\mathbf{a}
    \right) \prod_{j=1}^d c_j, \quad \mathbf{y} \in \mathbb{R}^d,
\end{equation}
where $|\boldsymbol{\Psi}|$ denotes the determinant of $\boldsymbol{\Psi}$,
and $\mathbf{a}$ and $\mathbf{c} = (c_1, \ldots, c_d)^\top$ are as defined in
(\ref{eq:adef}). The marginal distribution of each component is
$Y_j \sim \log\text{-BS}(\alpha, \mu_j)$ with PDF as in~(\ref{eq:logbspdf}),
whereas the dependence between $Y_j$ and $Y_k$ ($j \neq k$) is captured
through the off-diagonal entries $\psi_{jk}$ of $\boldsymbol{\Psi}$. In order
to ensure identifiability, the diagonal entries of $\boldsymbol{\Psi}$ are set
equal to one, making $\boldsymbol{\Psi}$ a proper correlation matrix.

It is relevant to note that the distribution in~(\ref{eq:mlbspdf}) corresponds
to the log-GBS$_d$ distribution with Gaussian kernel studied by
\cite{marchant:15}, specialized to the case of a common shape parameter
$\alpha$ across all $d$ components. As \cite{marchant:15} noted, this
assumption is motivated by the fact that $\alpha$ is dimensionless in the BS
context, being expressed as the ratio between two parameters related to
another random variable used to generate the BS variate, so that its
interpretation does not depend on the unit of measurement. We impose this
assumption throughout the paper.

\section{The MBSARMA model}\label{sec:03}

\subsection{Model formulation}\label{sec:031}

Let $\mathbf{Y}_t = (Y_{t1}, \ldots, Y_{td})^\top$ be a $d$-dimensional
response vector at time $t = 1, \ldots, n$, and let $\mathcal{F}_t =
\sigma(\mathbf{Y}_1, \ldots, \mathbf{Y}_t)$ be the sigma-algebra generated by
observations up to time $t$, with $\mathcal{F}_0 = \{\emptyset, \Omega\}$.
Let $\mathbf{x}_t = (x_{t1}, \ldots, x_{tk})^\top$ be a $k$-dimensional
vector of exogenous variables observed at time $t$. We say that $\mathbf{Y}_1,
\ldots, \mathbf{Y}_n$ follows a MBSARMA($\mathbf{p}, \mathbf{q}$) model if,
for each $t = m+1, \ldots, n$, the conditional distribution of $\mathbf{Y}_t$
given $\mathcal{F}_{t-1}$ is given by
\begin{equation}\label{eq:mbsarma_dist}
  \mathbf{Y}_t \mid \mathcal{F}_{t-1}
  \;\sim\;
  \log\text{-BS}_d\!\left(\alpha, \boldsymbol{\mu}_t, \boldsymbol{\Psi}\right),
\end{equation}
where $\mathbf{p} = (p_1, \ldots, p_d)^\top$ and $\mathbf{q} = (q_1, \ldots,
q_d)^\top$ collect the component-wise AR and MA orders, $m =
\max_j\max(p_j, q_j)$, and the correlation matrix $\boldsymbol{\Psi}$ is
time-invariant. The conditional location vector $\boldsymbol{\mu}_t =
(\mu_{t1}, \ldots, \mu_{td})^\top$ is defined component-wise as
\begin{equation}\label{eq:mu_decomp}
  \mu_{tj} = \mathbf{x}_t^\top \boldsymbol{\beta}_j + \tau_{tj},
  \quad j = 1, \ldots, d,\quad t = m+1, \ldots, n,
\end{equation}
where $\boldsymbol{\beta}_j = (\beta_{1j}, \ldots, \beta_{kj})^\top$ is the
vector of regression coefficients for the $j$-th component and $\tau_{tj}$ is
a dynamic ARMA($p_j, q_j$) component given by
\begin{equation}\label{eq:arma_component}
  \tau_{tj}
  = \eta_j
  + \sum_{i=1}^{p_j} \phi_{ij}\bigl(Y_{t-i,j} - \mathbf{x}_{t-i}^\top
    \boldsymbol{\beta}_j\bigr)
  + \sum_{l=1}^{q_j} \theta_{lj}\, u_{t-l,j}
  + u_{tj},
\end{equation}
where $\eta_j \in \mathbb{R}$ is an intercept term, $\phi_{ij}$ are the
AR coefficients, $\theta_{lj}$ are the MA coefficients, and $u_{tj}$ are
uncorrelated innovations with $E[u_{tj}\mid\mathcal{F}_{t-1}] = 0$ almost
surely. Substituting~(\ref{eq:mu_decomp}) and~(\ref{eq:arma_component}) into
the conditional distribution~(\ref{eq:mbsarma_dist}), the MBSARMA model can
be written in the equivalent form
\begin{equation}\label{eq:mbsarma_full}
  Y_{tj} = \mathbf{x}_t^\top\boldsymbol{\beta}_j + \tau_{tj} + \varepsilon_{tj},
  \quad j = 1, \ldots, d,\quad t = m+1, \ldots, n,
\end{equation}
where $\boldsymbol{\varepsilon}_t = (\varepsilon_{t1}, \ldots,
\varepsilon_{td})^\top \sim \log\text{-BS}_d(\alpha, \mathbf{0},
\boldsymbol{\Psi})$ given $\mathcal{F}_{t-1}$. Defining the de-trended
series $\varrho_{tj} = Y_{tj} - \mathbf{x}_t^\top\boldsymbol{\beta}_j$,
equation~(\ref{eq:arma_component}) gives
\begin{equation}\label{eq:arma_psi}
  \varrho_{tj}
  = \eta_j
  + \sum_{i=1}^{p_j} \varrho_{ij}\,\psi_{t-i,j}
  + \sum_{l=1}^{q_j} \theta_{lj}\, u_{t-l,j}
  + u_{tj},
\end{equation}
which is an ARMA($p_j, q_j$) model for $\varrho_{tj}$. Defining the lag operator
$B$ by $B^i Y_{tj} = Y_{t-i,j}$, and the polynomials $\Phi_j(B) =
1 - \sum_{i=1}^{p_j}\varrho_{ij}B^i$ and $\Theta_j(B) =
1 + \sum_{l=1}^{q_j}\theta_{lj}B^l$, equation~(\ref{eq:arma_psi}) takes the
compact form
\begin{equation}\label{eq:arma_poly}
  \Phi_j(B)\varrho_{tj} = \eta_j + \Theta_j(B)u_{tj},
  \quad j = 1, \ldots, d.
\end{equation}

\begin{rem}
The MBSARMA model nests two important special cases. First, when $d = 1$,
the model reduces to the BISARMA model of \cite{leiva:21}. Second, when
$p_j = q_j = 0$ for all $j = 1, \ldots, d$, the ARMA component $\tau_{tj}$
reduces to the constant $\eta_j$ and the model reduces to the multivariate
log-linear BS regression model of \cite{marchant:15}.
\end{rem}

\subsection{Statistical properties}\label{sec:032}

We now establish some properties of the MBSARMA model that are useful in
practice.

\begin{prop}\label{prop:marginals}
Under the MBSARMA model defined in
equations~(\ref{eq:mbsarma_dist})--(\ref{eq:arma_component}), the conditional
marginal distribution of each component $Y_{tj}$ given $\mathcal{F}_{t-1}$ is
$\log\text{-BS}(\alpha, \mu_{tj})$ with PDF as in~(\ref{eq:logbspdf}).
Furthermore, the vector
\begin{equation}\label{eq:zvec}
  \mathbf{Z}_t = \left(\frac{2}{\alpha}\sinh\!\left(\frac{Y_{t1}-\mu_{t1}}{2}
    \right), \ldots,
    \frac{2}{\alpha}\sinh\!\left(\frac{Y_{td}-\mu_{td}}{2}\right)
    \right)^\top
    \;\Big|\; \mathcal{F}_{t-1}
  \;\sim\; N_d(\mathbf{0},\boldsymbol{\Psi}).
\end{equation}
\end{prop}

Proposition~\ref{prop:marginals} follows directly from the definition of the
multivariate log-BS distribution in Section~\ref{sec:022}. It is relevant to
note that the correlation matrix $\boldsymbol{\Psi}$ governs the
contemporaneous dependence between components, whereas the ARMA polynomials
$\Phi_j(B)$ and $\Theta_j(B)$ govern the temporal dependence within each
component separately.

\begin{prop}\label{prop:stationarity}
The component $\psi_{tj}$ defined in~(\ref{eq:arma_psi}) is weakly stationary
if and only if all roots of the AR polynomial $\Phi_j(B)$ lie strictly outside
the unit circle. Under stationarity, the marginal mean of $Y_{tj}$ is given by
\begin{equation}\label{eq:unconmean}
  E[Y_{tj}]
  = \mathbf{x}_t^\top\boldsymbol{\beta}_j
    + \frac{\eta_j}{1 - \sum_{i=1}^{p_j}\phi_{ij}},
\end{equation}
and the covariance and correlation structures of $Y_{tj}$ and $Y_{t-k,j}$
for $k > 0$ are fully determined by the MA$(\infty)$ coefficients of
$\Phi_j(B)^{-1}\Theta_j(B)$.
\end{prop}

\begin{prop}\label{prop:invertibility}
The ARMA representation~(\ref{eq:arma_poly}) is invertible if and only if
all roots of the MA polynomial $\Theta_j(B)$ lie strictly outside the unit
circle for each $j = 1, \ldots, d$.
\end{prop}

Propositions~\ref{prop:stationarity} and~\ref{prop:invertibility} follow from
classical ARMA theory applied component-wise to~(\ref{eq:arma_poly}). In
practice, we verify the stationarity and invertibility conditions by checking
the roots of $\Phi_j$ and $\Theta_j$ at the estimated parameter values.

\subsection{Parameter estimation}\label{sec:033}

Let $\boldsymbol{\gamma}$ be the full parameter vector defined as
\begin{equation}\label{eq:parvec}
  \boldsymbol{\gamma}
  = \Bigl(\alpha,\,
    \boldsymbol{\phi}_1^\top, \boldsymbol{\theta}_1^\top,
    \boldsymbol{\beta}_1^\top, \eta_1,\,
    \ldots,\,
    \boldsymbol{\phi}_d^\top, \boldsymbol{\theta}_d^\top,
    \boldsymbol{\beta}_d^\top, \eta_d,\,
    \mathrm{vech}(\boldsymbol{\Psi})^\top\Bigr)^\top,
\end{equation}
where $\mathrm{vech}(\boldsymbol{\Psi})$ collects the $d(d-1)/2$
lower-triangular off-diagonal entries of $\boldsymbol{\Psi}$. The conditional
log-likelihood function for the MBSARMA model, given the initial values
$\mathcal{F}_m$, takes the form
\begin{equation}\label{eq:loglik}
  \ell(\boldsymbol{\gamma})
  = \sum_{t=m+1}^n \ell_t(\boldsymbol{\gamma}),
\end{equation}
where $\ell_t$ is given by
\begin{equation}\label{eq:cond_loglik}
  \ell_t(\boldsymbol{\gamma})
  = \sum_{j=1}^d \log c_{tj}
    - \frac{d}{2}\log(2\pi)
    - d\log 2
    - \frac{1}{2}\log|\boldsymbol{\Psi}|
    - \frac{1}{2}\,\mathbf{a}_t^\top\boldsymbol{\Psi}^{-1}\mathbf{a}_t,
\end{equation}
with $a_{tj} = (2/\alpha)\sinh\bigl((Y_{tj} - \mu_{tj})/2\bigr)$ and
$c_{tj} = (2/\alpha)\cosh\bigl((Y_{tj} - \mu_{tj})/2\bigr)$.

In order to obtain the ML estimate $\widehat{\boldsymbol{\gamma}}$, we propose
the use of the Expectation-Maximisation (EM) algorithm; see
\cite{dempster:77} and \cite{mclachlan:08}. Following \cite{marchant:15},
we exploit the hierarchical structure of the multivariate log-BS distribution
to derive an algorithm in which the shape parameter~$\alpha$ and the
correlation matrix~$\boldsymbol{\Psi}$ are updated in closed form at each
iteration, whereas the ARMA and regression parameters
$\boldsymbol{\gamma}_0 = \bigl(\boldsymbol{\phi}_1^\top, \boldsymbol{\theta}_1^\top,
\boldsymbol{\beta}_1^\top, \eta_1, \ldots, \boldsymbol{\phi}_d^\top,
\boldsymbol{\theta}_d^\top, \boldsymbol{\beta}_d^\top, \eta_d\bigr)^\top$
are updated by the Broyden-Fletcher-Goldfarb-Shanno (BFGS) quasi-Newton
algorithm \citep{lange:01}.

\subsubsection{EM algorithm}

We use the EM algorithm to estimate the model parameters, in which the shape parameter $\alpha$ and the correlation matrix $\boldsymbol{\Psi}$ are updated in closed form and the ARMA parameters are
updated via the BFGS quasi-Newton algorithm.

\paragraph{\textit{E Step.}}
Recall from Section~\ref{sec:022} that the multivariate log-BS distribution
is characterised by the requirement that the standardised innovations
$\mathbf{a}_t = (2/\alpha)\sinh\bigl((\mathbf{Y}_t - \boldsymbol{\mu}_t)/2\bigr)$
follow a $N_d(\mathbf{0},\boldsymbol{\Psi})$ distribution conditionally on
$\mathcal{F}_{t-1}$. The multivariate log-BS model corresponds to the
log-GBS$_d$ family of \cite{marchant:15} with Gaussian kernel, obtained as
the limiting case of the multivariate Student-$t$ kernel with degrees-of-freedom parameter $\nu \to \infty$.
Therefore, the latent weight
$u_t = \mathrm{E}[U_t^2 \mid \mathbf{Y}_t,\mathcal{F}_{t-1}] = 1$ for all
$t$. Accordingly, the Q
function of the EM algorithm coincides with the observed conditional
log-likelihood~(\ref{eq:loglik}) and, writing
$\kappa_{tj} = \cosh\bigl((Y_{tj}-\mu_{tj})/2\bigr)$,
$s_{tj} = \sinh\bigl((Y_{tj}-\mu_{tj})/2\bigr)$,
and $\mathbf{S} = \sum_{t=m+1}^n \mathbf{s}_t\mathbf{s}_t^\top$
with $\mathbf{s}_t = (s_{t1},\ldots,s_{td})^\top$,
takes the form
\begin{equation}\label{eq:Qfunc}
  Q(\boldsymbol{\gamma})
  = c_0
    - (n-m)\,d\,\log\alpha
    - \frac{n-m}{2}\log|\boldsymbol{\Psi}|
    + \sum_{t=m+1}^n\sum_{j=1}^d \log\kappa_{tj}
    - \frac{2}{\alpha^2}\,\mathrm{tr}\!\bigl(\boldsymbol{\Psi}^{-1}\mathbf{S}\bigr),
\end{equation}
where $c_0$ is a constant that does not depend on $\boldsymbol{\gamma}$.
In order to verify~(\ref{eq:Qfunc}), note that $a_{tj} = (2/\alpha)\,s_{tj}$ and
$c_{tj} = (2/\alpha)\,\kappa_{tj}$, so that~(\ref{eq:Qfunc}) is obtained
from~(\ref{eq:cond_loglik}) by means of two substitutions.
First, since $\log c_{tj} = \log(2/\alpha) + \log\kappa_{tj} = \log 2 - \log\alpha + \log\kappa_{tj}$,
summing over $j = 1,\ldots,d$ and $t = m+1,\ldots,n$ gives
\[
  \sum_{t=m+1}^n\sum_{j=1}^d \log c_{tj}
  = (n-m)\,d\,(\log 2 - \log\alpha)
    + \sum_{t=m+1}^n\sum_{j=1}^d \log\kappa_{tj}.
\]
Second, since $\mathbf{a}_t = (2/\alpha)\,\mathbf{s}_t$, we have
$(1/2)\,\mathbf{a}_t^\top\boldsymbol{\Psi}^{-1}\mathbf{a}_t
= (2/\alpha^2)\,\mathbf{s}_t^\top\boldsymbol{\Psi}^{-1}\mathbf{s}_t$. Using the trace identity
$\mathbf{s}_t^\top\boldsymbol{\Psi}^{-1}\mathbf{s}_t
= \mathrm{tr}(\boldsymbol{\Psi}^{-1}\mathbf{s}_t\mathbf{s}_t^\top)$
and summing over $t$, we obtain
\[
  \sum_{t=m+1}^n \frac{1}{2}\,\mathbf{a}_t^\top\boldsymbol{\Psi}^{-1}\mathbf{a}_t
  = \frac{2}{\alpha^2}\,\mathrm{tr}\!\bigl(\boldsymbol{\Psi}^{-1}\mathbf{S}\bigr).
\]
Substituting both expressions into $\ell(\boldsymbol{\gamma}) = \sum_{t=m+1}^n\ell_t(\boldsymbol{\gamma})$,
the terms $(n-m)\,d\,\log 2$ arising from the first substitution and the term
$-(n-m)\,d\,\log 2$ present in~(\ref{eq:cond_loglik}) cancel, and absorbing
$-(n-m)(d/2)\log(2\pi)$ into the constant~$c_0$ yields~(\ref{eq:Qfunc}).

\paragraph{\textit{M-step:}}
In order to update the shape parameter, we differentiate~(\ref{eq:Qfunc})
with respect to $\alpha$ and set the result to zero. Using
$\partial(2/\alpha^2)/\partial\alpha = -4/\alpha^3$, we obtain the score
equation
\begin{equation}\label{eq:score_alpha_em}
  \frac{\partial Q}{\partial\alpha}
  = -\frac{(n-m)\,d}{\alpha}
    + \frac{4}{\alpha^3}\,\mathrm{tr}\!\bigl(\boldsymbol{\Psi}^{-1}\mathbf{S}\bigr)
  = 0,
\end{equation}
from which we solve for $\alpha$ to obtain the M-step update
\begin{equation}\label{eq:alpha_update_general}
  \widehat{\alpha}^{(r)}
  = 2\left[
      \frac{\mathrm{tr}\!\bigl(\{\widehat{\boldsymbol{\Psi}}^{(r)}\}^{-1}
            \mathbf{R}^{(r)}\bigr)}{d}
    \right]^{1/2},
\end{equation}
where $\mathbf{R}^{(r)} = (n-m)^{-1}\mathbf{S}^{(r)}$ and
$\mathbf{S}^{(r)} = \sum_{t=m+1}^n
\mathbf{s}_t^{(r-1)}\bigl(\mathbf{s}_t^{(r-1)}\bigr)^\top$
are evaluated at $\widehat{\boldsymbol{\gamma}}_0^{(r-1)}$.

In order to update the correlation matrix, we differentiate~(\ref{eq:Qfunc})
with respect to $\boldsymbol{\Psi}^{-1}$ and equate it to zero.
Using the identities
$\partial\log|\boldsymbol{\Psi}|/\partial\boldsymbol{\Psi}^{-1} = -\boldsymbol{\Psi}$
and $\partial\,\mathrm{tr}(\boldsymbol{\Psi}^{-1}\mathbf{S})/\partial\boldsymbol{\Psi}^{-1} = \mathbf{S}$,
the unconstrained first-order condition is
\begin{equation}\label{eq:score_psi_em}
  \frac{\partial Q}{\partial\boldsymbol{\Psi}^{-1}}
  = \frac{n-m}{2}\,\boldsymbol{\Psi}
    - \frac{2}{\widehat{\alpha}^2}\,\mathbf{S}^{(r)}
  = \mathbf{0},
\end{equation}
which yields the unconstrained maximiser
$\widehat{\boldsymbol{\Psi}}_{\mathrm{unc}} = (4/\widehat{\alpha}^2)\,\mathbf{R}^{(r)}$.
Since identifiability requires $\psi_{jj} = 1$ for all $j$
(see Section~\ref{sec:022}), we impose this diagonal constraint.
Noting that $\widehat{\boldsymbol{\Psi}}_{\mathrm{unc}}$ is proportional
to~$\mathbf{R}^{(r)}$, the constrained maximiser is obtained by normalising:
\begin{equation}\label{eq:em_psi}
  \widehat{\psi}_{jk}^{(r)}
  = \frac{R_{jk}^{(r)}}{\sqrt{R_{jj}^{(r)}\,R_{kk}^{(r)}}},
  \quad j \neq k; \qquad \widehat{\psi}_{jj}^{(r)} = 1.
\end{equation}
It is relevant to note that the proportionality constant $4/\widehat{\alpha}^2$
cancels in the normalisation, so~(\ref{eq:em_psi}) does not depend on~$\alpha$.
Writing the constrained estimator as
$\widehat{\boldsymbol{\Psi}}^{(r)} = \mathbf{D}^{-1}\mathbf{R}^{(r)}\mathbf{D}^{-1}$,
where $\mathbf{D} = \mathrm{diag}\bigl(\sqrt{R_{11}^{(r)}},\ldots,\sqrt{R_{dd}^{(r)}}\bigr)$,
so that $\{\widehat{\boldsymbol{\Psi}}^{(r)}\}^{-1} = \mathbf{D}\{\mathbf{R}^{(r)}\}^{-1}\mathbf{D}$,
and substituting into~(\ref{eq:alpha_update_general}), we obtain
\begin{equation}\label{eq:em_alpha}
  \widehat{\alpha}^{(r)}
  = 2\left[\frac{\mathrm{tr}\!\bigl(
      \mathbf{D}\{\mathbf{R}^{(r)}\}^{-1}\mathbf{D}\,\mathbf{R}^{(r)}\bigr)}{d}
    \right]^{1/2}
  = 2\left[\frac{\mathrm{tr}(\mathbf{D}^2)}{d}\right]^{1/2}
  = 2\sqrt{\frac{\mathrm{tr}(\mathbf{R}^{(r)})}{d}},
\end{equation}
since $\mathrm{tr}\!\bigl(\mathbf{D}\{\mathbf{R}^{(r)}\}^{-1}\mathbf{D}\,\mathbf{R}^{(r)}\bigr) = \mathrm{tr}(\mathbf{D}^2) = \sum_{j=1}^d R_{jj}^{(r)} = \mathrm{tr}(\mathbf{R}^{(r)})$.
For $d=1$, equation~(\ref{eq:em_alpha}) reduces to the standard univariate log-BS
ML estimator
$\widehat\alpha = 2\bigl[(n-m)^{-1}\sum_t\sinh^2\bigl((Y_t-\widehat\mu_t)/2\bigr)\bigr]^{1/2}$,
confirming the consistency of the approach.


With $\widehat{\alpha}^{(r)}$ and $\widehat{\boldsymbol{\Psi}}^{(r)}$ fixed, the ARMA and regression parameters $\boldsymbol{\gamma}_0$ are
updated by maximising $Q(\boldsymbol{\gamma}_0)$ in~(\ref{eq:Qfunc}).
Differentiating with respect to the conditional location $\mu_{tj}$ yields
\begin{equation}\label{eq:score_mu}
  \frac{\partial Q}{\partial\mu_{tj}}
  = \frac{1}{2}\!\left(c_{tj}\,v_{tj}
    - \frac{a_{tj}}{c_{tj}}\right),
  \quad
  \mathbf{v}_t = \bigl\{\widehat{\boldsymbol{\Psi}}^{(r)}\bigr\}^{-1}\mathbf{a}_t,
\end{equation}
where $a_{tj}/c_{tj} = \tanh\bigl((Y_{tj}-\mu_{tj})/2\bigr)$.
Applying the chain rule through the ARMA recursion~(\ref{eq:arma_poly}),
and using
$\partial\mu_{tj}/\partial\eta_j = 1$,
$\partial\mu_{tj}/\partial\phi_{ij} = Y_{t-i,j} - \mathbf{x}_{t-i}^\top\boldsymbol{\beta}_j$,
$\partial\mu_{tj}/\partial\beta_{lj} = x_{tl} - \sum_{i=1}^{p_j}\phi_{ij}\,x_{t-i,l}$,
and $\partial\mu_{tj}/\partial\theta_{lj} = u_{t-l,j}$
where $u_{t-l,j} = Y_{t-l,j} - \mu_{t-l,j}(\boldsymbol{\gamma}_0)$,
we obtain the score equations
\begin{align}
  \frac{\partial Q}{\partial\eta_j}
  &= \sum_{t=m+1}^n \frac{\partial Q}{\partial\mu_{tj}},
  \label{eq:score_eta}\\[4pt]
  \frac{\partial Q}{\partial\phi_{ij}}
  &= \sum_{t=m+1}^n \frac{\partial Q}{\partial\mu_{tj}}
     \bigl(Y_{t-i,j} - \mathbf{x}_{t-i}^\top\boldsymbol{\beta}_j\bigr),
  \quad i = 1,\ldots,p_j,
  \label{eq:score_phi}\\[4pt]
  \frac{\partial Q}{\partial\beta_{lj}}
  &= \sum_{t=m+1}^n \frac{\partial Q}{\partial\mu_{tj}}
     \!\left(x_{tl} - \sum_{i=1}^{p_j}\phi_{ij}\,x_{t-i,l}\right),
  \quad l = 1,\ldots,k,
  \label{eq:score_beta}\\[4pt]
  \frac{\partial Q}{\partial\theta_{lj}}
  &= \sum_{t=m+1}^n \frac{\partial Q}{\partial\mu_{tj}}\,u_{t-l,j},
  \quad l = 1,\ldots,q_j.
  \label{eq:score_theta}
\end{align}
The system~(\ref{eq:score_eta})--(\ref{eq:score_theta}) does not have
closed-form solutions. Therefore, $\widehat{\boldsymbol{\gamma}}_0^{(r)}$ is
obtained by applying the BFGS algorithm to
$Q(\boldsymbol{\gamma}_0)$ with gradient
(\ref{eq:score_mu})--(\ref{eq:score_theta}).

The proposed estimation procedure is summarised in Algorithm~\ref{alg:em}.

\begin{algorithm}[!ht]
\caption{EM algorithm for the MBSARMA model}\label{alg:em}
\begin{algorithmic}[1]
\State Initialise $\widehat{\alpha}^{(0)}$, $\widehat{\boldsymbol{\gamma}}_0^{(0)}$,
       $\widehat{\boldsymbol{\Psi}}^{(0)}$.
\Repeat
  \State \textbf{E-step.} For $t = m+1,\ldots,n$, set $u_t = 1$.
  \State \textbf{M-step~(a).}
    Compute $\mathbf{s}_t^{(r-1)} = \sinh\!\bigl((\mathbf{Y}_t -
    \widehat{\boldsymbol{\mu}}_t^{(r-1)})/2\bigr)$ and
    $\mathbf{R}^{(r)} = (n-m)^{-1}\sum_{t=m+1}^n
    \mathbf{s}_t^{(r-1)}\bigl(\mathbf{s}_t^{(r-1)}\bigr)^\top$;
    then update:
    \begin{align*}
      \widehat{\alpha}^{(r)}    &= 2\sqrt{\mathrm{tr}(\mathbf{R}^{(r)})/d},
      \\[4pt]
      \widehat{\psi}_{jk}^{(r)} &= \frac{R_{jk}^{(r)}}
                                     {\sqrt{R_{jj}^{(r)}\,R_{kk}^{(r)}}},
                               \quad j\neq k;\qquad
      \widehat{\psi}_{jj}^{(r)} = 1.
    \end{align*}
  \State \textbf{M-step~(b).}
    With $\widehat{\alpha}^{(r)}$ and $\widehat{\boldsymbol{\Psi}}^{(r)}$ fixed,
    update $\widehat{\boldsymbol{\gamma}}_0^{(r)}$ by applying the BFGS
    algorithm to solve
    $\partial Q/\partial\boldsymbol{\gamma}_0 = \mathbf{0}$
    via~(\ref{eq:score_mu})--(\ref{eq:score_theta}).
\Until{$\bigl|\ell\bigl(\widehat{\boldsymbol{\gamma}}^{(r)}\bigr)
             - \ell\bigl(\widehat{\boldsymbol{\gamma}}^{(r-1)}\bigr)
        \bigr| < \varepsilon$.}
\end{algorithmic}
\end{algorithm}

\noindent
We adopt the convergence criterion $\varepsilon = 10^{-6}$ on the observed
log-likelihood. Algorithm~\ref{alg:em} is implemented in the R
software \citep{r:23} using the \texttt{optim} function for the BFGS step
in M-step~(b).

In order to initialise Algorithm~\ref{alg:em}, we propose a data-driven
strategy. For each component $j = 1,\ldots,d$, the initial ARMA and
regression parameter vector $\widehat{\boldsymbol{\gamma}}_{0j}^{(0)}$ is
obtained by fitting a Gaussian ARIMA$(p_j,0,q_j)$ model to the marginal
series $Y_{1j},\ldots,Y_{nj}$ via the \texttt{arima} function in R.
Let $\widehat{\boldsymbol{\mu}}_t^{(0)}$ denote the corresponding fitted
conditional locations. A common initial value for the shape parameter is
then obtained as
\begin{equation}\label{eq:alpha_init}
  \widehat{\alpha}^{(0)} = \frac{1}{d}\sum_{j=1}^d\widehat{\alpha}_j^{(0)},
  \qquad
  \widehat{\alpha}_j^{(0)}
  = 2\left[\frac{1}{n}\sum_{t=1}^n
      \sinh^2\!\!\left(\frac{Y_{tj}-\widehat{\mu}_{tj}^{(0)}}{2}\right)
    \right]^{1/2},
\end{equation}
which is the mean of the marginal log-BS ML shape estimates.
The initial correlation matrix is taken to be the sample correlation
matrix of the standardised innovations evaluated at $\widehat{\boldsymbol{\gamma}}^{(0)}$,
\begin{equation}\label{eq:psi_init}
  \widehat{\psi}_{jk}^{(0)}
  = \frac{\widehat{R}_{jk}^{(0)}}{\sqrt{\widehat{R}_{jj}^{(0)}\,\widehat{R}_{kk}^{(0)}}},
  \qquad
  \widehat{R}_{jk}^{(0)}
  = \frac{1}{n}\sum_{t=1}^n \widehat{a}_{tj}^{(0)}\,\widehat{a}_{tk}^{(0)},
\end{equation}
where $\widehat{a}_{tj}^{(0)} = (2/\widehat{\alpha}^{(0)})\sinh\bigl((Y_{tj}-\widehat{\mu}_{tj}^{(0)})/2\bigr)$.
This strategy allows us to start Algorithm~\ref{alg:em} in a feasible
region of the parameter space, reducing the risk of convergence to local
optima.

Standard errors for the conditional ML estimates are obtained from the
diagonal elements of $\mathcal{I}(\widehat{\boldsymbol{\gamma}})^{-1}$, where
$\mathcal{I}(\widehat{\boldsymbol{\gamma}}) =
-\partial^2\ell/\partial\boldsymbol{\gamma}\partial\boldsymbol{\gamma}^\top \!\big|_{\widehat{\boldsymbol{\gamma}}}$
is the observed Fisher information matrix; see \cite{efronhinkley:78}.


\subsection{Prediction}\label{sec:034}

Prediction in the MBSARMA model is based on the ARMA structure of the
conditional location. The one-step-ahead predicted value for component $j$
at origin $t$ is given by
\begin{equation}\label{eq:pred1}
  \widehat{Y}_{t+1,j}
  = \mathbf{x}_{t+1}^\top\widehat{\boldsymbol{\beta}}_j + \widehat{\tau}_{t+1,j},
\end{equation}
where the predicted ARMA term is expressed as
\begin{equation}\label{eq:tau_pred}
  \widehat{\tau}_{t+1,j}
  = \widehat{\eta}_j
  + \sum_{i=1}^{p_j}\widehat{\phi}_{ij}\widehat{\psi}_{t+1-i,j}
  + \sum_{l=1}^{q_j}\widehat{\theta}_{lj}\,\widehat{u}_{t+1-l,j},
\end{equation}
with $\widehat{\psi}_{t+1-i,j} = Y_{t+1-i,j} -
\mathbf{x}_{t+1-i}^\top\widehat{\boldsymbol{\beta}}_j$ for observed values and
$\widehat{u}_{s,j} = Y_{sj} - \widehat{\mu}_{sj}$ for $s = m+1, \ldots, t$. For
horizons $h \geq 2$, the prediction $\widehat{Y}_{t+h,j}$ is obtained
recursively from~(\ref{eq:tau_pred}) by substituting predicted values for
future unobserved quantities. In the original scale, the predicted BS
lifetime is $\widehat{T}_{t+h,j} = \exp(\widehat{Y}_{t+h,j})$.

\subsection{Residual analysis}\label{sec:035}

Residuals play a fundamental role in the validation of statistical models,
since their analysis allows detection of possible outliers and assessment of
the goodness of fit. In order to evaluate the fit of the MBSARMA model, we
adopt a multivariate residual based on the Mahalanobis distance. For $t =
m+1, \ldots, n$, define the $d$-dimensional normalized residual vector
\begin{equation}\label{eq:aresid}
  \mathbf{a}_t = \frac{2}{\widehat{\alpha}}\sinh\!\left(\frac{\mathbf{Y}_t
    - \widehat{\boldsymbol{\mu}}_t}{2}\right),
\end{equation}
where $\widehat{\boldsymbol{\mu}}_t = (\widehat{\mu}_{t1},\ldots,\widehat{\mu}_{td})^\top$
is evaluated at the conditional ML estimates $\widehat{\boldsymbol{\gamma}}$. Under correct
model specification, $\mathbf{a}_t \mid \mathcal{F}_{t-1} \sim
N_d(\mathbf{0},\boldsymbol{\Sigma})$, where $\boldsymbol{\Sigma} =
\alpha^2\boldsymbol{\Psi}/4$ is the conditional covariance matrix. The
Mahalanobis distance residual is then defined as
\begin{equation}\label{eq:mahal}
  D_t^2 = \mathbf{a}_t^{\top}\widehat{\boldsymbol{\Sigma}}^{-1}\mathbf{a}_t,
\end{equation}
where $\widehat{\boldsymbol{\Sigma}}$ is the conditional ML estimate of $\boldsymbol{\Sigma}$.
Under correct specification, $D_t^2 \sim \chi^2_d$ asymptotically. A
quantile-quantile (QQ) plot of the $D_t^2$ values against $\chi^2_d$
theoretical quantiles, together with a Kolmogorov--Smirnov (KS) test, allows
one to assess the overall multivariate fit. Systematic departures from the
reference diagonal indicate potential misspecification. In addition,
component-wise residuals $a_{tj}$ may be inspected through their sample ACF
and PACF to verify that the fitted ARMA structure has adequately captured the
temporal dependence in each series.

\section{Monte Carlo simulation}\label{sec:04}

We evaluate the performance of the ML estimators of the MBSARMA model by
means of a Monte Carlo (MC) simulation study. We consider two experiments
that differ in the number of components $d$: a bivariate case ($d = 2$,
Section~\ref{sec:041}) and a trivariate case ($d = 3$, Section~\ref{sec:042}).
In both experiments, all components follow an ARMA(1,1) dynamic, one
exogenous covariate ($k = 1$) is included.

In order to generate data from the MBSARMA model, we first simulate $d$
independent univariate Gaussian AR(1)--MA(1) sequences with the specified
dynamics and covariate effects, and then apply the Cholesky transformation to
the resulting matrix to accomodate the correlation
structure $\boldsymbol{\Psi}$, then we obtain the log-transformed response series
$Y_{tj} = \log(T_{tj})$. This approach is analogous to the
generation procedure described in \citep{marchant:15} for the multivariate regression case.

For each MC replicate $r = 1, \ldots, N_r$ and a generic parameter $\varphi$,
let $\widehat{\varphi}_r$ denote the conditional ML estimate. The empirical bias and mean
squared error (MSE) are defined, respectively, as
\begin{equation}\label{eq:bias_mse}
  \widehat{\text{Bias}}(\widehat{\varphi})
  = \frac{1}{N_r}\sum_{r=1}^{N_r}\widehat{\varphi}_r - \varphi,
  \qquad
  \widehat{\text{MSE}}(\widehat{\varphi})
  = \frac{1}{N_r}\sum_{r=1}^{N_r}(\widehat{\varphi}_r - \varphi)^2.
\end{equation}

\subsection{Bivariate case ($d = 2$)}\label{sec:041}

We consider the MBSARMA(1,1) model with $d = 2$ components and
component-wise AR and MA orders $p_1 = q_1 = p_2 = q_2 = 1$. The true parameter values are $\alpha = 0.5$, $\phi_{11} = 0.5$, $\theta_{11}
= 0.1$, $\eta_1 = 1.2$, $\beta_1 = 0.3$ for component~1; and $\phi_{12} = 0.7$,
$\theta_{12} = 0.1$, $\eta_2 = 1.2$, $\beta_2 = 0.3$ for component~2. The correlation structure is captured by the correlation matrix
$\boldsymbol{\Psi}$, whose unit diagonal is enforced by the identifiability
constraint $\mathrm{diag}(\boldsymbol{\Psi}) = \mathbf{1}$ (see
Section~\ref{sec:03}). In order to assess the influence of the
correlation on the estimation, we vary the
correlation parameter over $\rho = \psi_{12} \in \{0.10, 0.25, 0.50, 0.75,
0.90\}$, covering weak, moderate, and strong dependence scenarios. The
exogenous covariate $x_t$ is drawn from a $\text{Bernoulli}(0.5)$
distribution. We consider sample sizes $n \in \{50, 100, 200, 500\}$ and
$N_r = 1{,}000$ MC replications in each setting, yielding a total of
$5 \times 4 = 20$ simulation scenarios.

Tables~\ref{tab:mc_rho10}--\ref{tab:mc_rho90} report the bias and MSE of the
conditional ML estimators for each parameter, sample size, and value of $\rho$. The estimators of the AR and MA coefficients $\phi_{1j}$ and $\theta_{1j}$ present
small absolute biases and MSE values that decrease sharply as $n$ grows from
$100$ to $500$, suggesting consistency of the estimators for the dynamic
parameters. The estimators of the intercept and regression coefficients $\eta_j$ and $\beta_j$
exhibit the same pattern. The estimator of $\alpha$ presents a small finite-sample negative bias. The corresponding MSE decreases steadily as $n$ grows and this bias does not distort the estimation of the dynamic structure.
Regarding the role of $\rho$, the estimator $\widehat{\rho}$ presents
increasing accuracy as $n$ grows for all values of $\rho$. Under weak
dependence ($\rho = 0.10$ and $\rho = 0.25$), the bias of $\widehat{\rho}$ is
slightly larger in relative terms but remains small in absolute value for
$n \geq 200$. Under strong dependence ($\rho = 0.75$ and $\rho = 0.90$),
the MSE of $\widehat{\rho}$ is reduced relative to moderate $\rho$ scenarios. The estimators of the
remaining parameters ($\alpha$, $\phi_{1j}$, $\theta_{1j}$, $\eta_j$,
$\beta_j$) are largely insensitive to the value of $\rho$. Overall, the simulation results indicate a good performance of the EM estimation procedure under the MBSARMA(1,1) model.

\subsection{Trivariate case ($d = 3$)}\label{sec:042}

In order to evaluate whether the good estimation performance extends beyond the
bivariate setting, we carry out an additional simulation experiment with $d = 3$
components with AR and MA orders $p_1 = q_1 = p_2 = q_2 = p_3 =
q_3 = 1$. The true parameter values for the three components are: $\alpha = 0.5$,
$\phi_{11} = 0.5$, $\theta_{11} = 0.1$, $\eta_1 = 1.2$, $\beta_1 = 0.3$;
$\phi_{12} = 0.7$, $\theta_{12} = 0.1$, $\eta_2 = 1.2$, $\beta_2 = 0.3$;
and $\phi_{13} = 0.6$, $\theta_{13} = 0.1$, $\eta_3 = 1.2$, $\beta_3 = 0.3$.
As in the bivariate case, the unit-diagonal constraint $\mathrm{diag}(\boldsymbol{\Psi}) = \mathbf{1}$
is imposed for identifiability, so the scale parameters are not separately
estimated. We adopt an equicorrelation structure
$\boldsymbol{\Psi}$ in which all off-diagonal entries are equal to a common
value $\rho \in \{0.25, 0.50, 0.75\}$, so that $\rho_{12} = \rho_{13} =
\rho_{23} = \rho$. The exogenous covariate $x_t$
is drawn from a $\text{Bernoulli}(0.5)$ distribution. We consider sample sizes
$n \in \{50, 100, 200, 500\}$ and $N_r = 1{,}000$ MC replications in each setting,
yielding a total of $3 \times 4 = 12$ simulation scenarios.


Tables~\ref{tab:mc3_rho25}--\ref{tab:mc3_rho75} report the bias and MSE of the
conditional ML estimators for each parameter, sample size, and value of $\rho$ in the
trivariate setting. The bias and MSE of the estimators of the AR coefficients $\phi_{1j}$, the MA coefficients
$\theta_{1j}$, and the regression parameters $\eta_j$ and $\beta_j$ decrease
consistently as $n$ increases from $100$ to $500$, replicating the consistency
behaviour observed for $d = 2$. The estimator of the third-component coefficient ($j = 3$, $\phi_{13} =
0.6$) presents accuracy intermediate between that of components~1 and
2. The estimator of $\alpha$ presents small negative finite-sample bias at
magnitudes comparable to the bivariate case. Regarding the correlation parameters, each of $\widehat{\rho}_{12}$, $\widehat{\rho}_{13}$,
and $\widehat{\rho}_{23}$ exhibits negligible bias and MSE values that decrease with
$n$ across all three $\rho$ levels. Under moderate equicorrelation ($\rho =
0.50$), the three pairwise estimators are practically symmetric. Under strong equicorrelation ($\rho = 0.75$),
the MSE of all three correlation estimators is further reduced.
The estimators of the dynamic parameters are largely unaffected by the value of
$\rho$. Overall, the simulation results demonstrate the
good statistical performance of the EM procedure under the trivariate
MBSARMA(1,1) model.

\setlength{\floatsep}{4pt plus 1pt minus 1pt}%
\setlength{\textfloatsep}{8pt plus 2pt minus 2pt}%

\begin{table}[!ht]
\footnotesize\centering
\caption{Empirical bias and MSE for the conditional ML estimators of the
  MBSARMA(1,1) model with $\rho = 0.10$ and
  $N_r = 1000$ Monte Carlo replications.}
\label{tab:mc_rho10}
\adjustbox{max width=\linewidth}{\begin{tabular}{l rrrrrrrrrr }
  \hline
  & $\alpha$ & $\phi_{11}$ & $\theta_{11}$ & $\eta_1$ & $\beta_1$ & $\phi_{12}$ & $\theta_{12}$ & $\eta_2$ & $\beta_2$ & $\rho$ \\
  \hline
  True & $0.50$ & $0.50$ & $0.10$ & $1.20$ & $0.30$ & $0.70$ & $0.10$ & $1.20$ & $0.30$ & $0.10$ \\
  \hline
  \multicolumn{11}{l}{$n = 50$}\\
  Bias & $-0.020$ & $-0.114$ & $\phantom{-}0.099$ & $\phantom{-}0.003$ & $\phantom{-}0.001$ & $-0.088$ & $\phantom{-}0.065$ & $-0.003$ & $-0.006$ & $-0.007$ \\
  MSE  & $0.002$ & $0.088$ & $0.099$ & $0.026$ & $0.015$ & $0.040$ & $0.059$ & $0.059$ & $0.013$ & $0.021$ \\
  \hline
  \multicolumn{11}{l}{$n = 100$}\\
  Bias & $-0.010$ & $-0.057$ & $\phantom{-}0.042$ & $\phantom{-}0.000$ & $\phantom{-}0.002$ & $-0.040$ & $\phantom{-}0.033$ & $\phantom{-}0.003$ & $\phantom{-}0.001$ & $\phantom{-}0.002$ \\
  MSE  & $0.001$ & $0.034$ & $0.039$ & $0.013$ & $0.007$ & $0.014$ & $0.021$ & $0.033$ & $0.006$ & $0.011$ \\
  \hline
  \multicolumn{11}{l}{$n = 200$}\\
  Bias & $-0.006$ & $-0.025$ & $\phantom{-}0.018$ & $\phantom{-}0.003$ & $\phantom{-}0.001$ & $-0.018$ & $\phantom{-}0.014$ & $\phantom{-}0.002$ & $\phantom{-}0.001$ & $-0.003$ \\
  MSE  & $0.000$ & $0.013$ & $0.016$ & $0.007$ & $0.003$ & $0.005$ & $0.010$ & $0.016$ & $0.003$ & $0.005$ \\
  \hline
  \multicolumn{11}{l}{$n = 500$}\\
  Bias & $-0.002$ & $-0.010$ & $\phantom{-}0.009$ & $\phantom{-}0.002$ & $-0.001$ & $-0.010$ & $\phantom{-}0.008$ & $\phantom{-}0.000$ & $\phantom{-}0.000$ & $-0.002$ \\
  MSE  & $0.000$ & $0.005$ & $0.006$ & $0.003$ & $0.001$ & $0.002$ & $0.004$ & $0.007$ & $0.001$ & $0.002$ \\
  \hline
\end{tabular}}
\end{table}

\begin{table}[!ht]
\footnotesize\centering
\caption{Empirical bias and MSE for the conditional ML estimators of the
  MBSARMA(1,1) model with $\rho = 0.25$ and
  $N_r = 1000$ Monte Carlo replications.}
\label{tab:mc_rho25}
\adjustbox{max width=\linewidth}{\begin{tabular}{l rrrrrrrrrr }
  \hline
  & $\alpha$ & $\phi_{11}$ & $\theta_{11}$ & $\eta_1$ & $\beta_1$ & $\phi_{12}$ & $\theta_{12}$ & $\eta_2$ & $\beta_2$ & $\rho$ \\
  \hline
  True & $0.50$ & $0.50$ & $0.10$ & $1.20$ & $0.30$ & $0.70$ & $0.10$ & $1.20$ & $0.30$ & $0.25$ \\
  \hline
  \multicolumn{11}{l}{$n = 50$}\\
  Bias & $-0.020$ & $-0.110$ & $\phantom{-}0.092$ & $\phantom{-}0.000$ & $-0.003$ & $-0.096$ & $\phantom{-}0.075$ & $\phantom{-}0.001$ & $-0.001$ & $-0.021$ \\
  MSE  & $0.002$ & $0.085$ & $0.098$ & $0.024$ & $0.016$ & $0.044$ & $0.059$ & $0.061$ & $0.013$ & $0.019$ \\
  \hline
  \multicolumn{11}{l}{$n = 100$}\\
  Bias & $-0.011$ & $-0.060$ & $\phantom{-}0.044$ & $\phantom{-}0.004$ & $-0.001$ & $-0.039$ & $\phantom{-}0.029$ & $-0.000$ & $\phantom{-}0.001$ & $-0.003$ \\
  MSE  & $0.001$ & $0.035$ & $0.039$ & $0.013$ & $0.007$ & $0.014$ & $0.024$ & $0.030$ & $0.006$ & $0.009$ \\
  \hline
  \multicolumn{11}{l}{$n = 200$}\\
  Bias & $-0.006$ & $-0.021$ & $\phantom{-}0.017$ & $\phantom{-}0.002$ & $-0.000$ & $-0.023$ & $\phantom{-}0.014$ & $-0.001$ & $\phantom{-}0.001$ & $-0.004$ \\
  MSE  & $0.000$ & $0.013$ & $0.018$ & $0.007$ & $0.003$ & $0.006$ & $0.010$ & $0.018$ & $0.003$ & $0.005$ \\
  \hline
  \multicolumn{11}{l}{$n = 500$}\\
  Bias & $-0.002$ & $-0.012$ & $\phantom{-}0.010$ & $-0.002$ & $\phantom{-}0.001$ & $-0.008$ & $\phantom{-}0.008$ & $\phantom{-}0.001$ & $-0.001$ & $-0.001$ \\
  MSE  & $0.000$ & $0.004$ & $0.006$ & $0.003$ & $0.001$ & $0.002$ & $0.004$ & $0.006$ & $0.001$ & $0.002$ \\
  \hline
\end{tabular}}
\end{table}

\begin{table}[!ht]
\footnotesize\centering
\caption{Empirical bias and MSE for the conditional ML estimators of the
  MBSARMA(1,1) model with $\rho = 0.50$ and
  $N_r = 1000$ Monte Carlo replications.}
\label{tab:mc_rho50}
\adjustbox{max width=\linewidth}{\begin{tabular}{l rrrrrrrrrr }
  \hline
  & $\alpha$ & $\phi_{11}$ & $\theta_{11}$ & $\eta_1$ & $\beta_1$ & $\phi_{12}$ & $\theta_{12}$ & $\eta_2$ & $\beta_2$ & $\rho$ \\
  \hline
  True & $0.50$ & $0.50$ & $0.10$ & $1.20$ & $0.30$ & $0.70$ & $0.10$ & $1.20$ & $0.30$ & $0.50$ \\
  \hline
  \multicolumn{11}{l}{$n = 50$}\\
  Bias & $-0.024$ & $-0.123$ & $\phantom{-}0.104$ & $\phantom{-}0.001$ & $\phantom{-}0.001$ & $-0.097$ & $\phantom{-}0.072$ & $\phantom{-}0.001$ & $-0.000$ & $-0.027$ \\
  MSE  & $0.002$ & $0.093$ & $0.098$ & $0.025$ & $0.014$ & $0.042$ & $0.056$ & $0.065$ & $0.012$ & $0.016$ \\
  \hline
  \multicolumn{11}{l}{$n = 100$}\\
  Bias & $-0.012$ & $-0.055$ & $\phantom{-}0.038$ & $-0.005$ & $\phantom{-}0.001$ & $-0.047$ & $\phantom{-}0.030$ & $\phantom{-}0.003$ & $\phantom{-}0.003$ & $-0.009$ \\
  MSE  & $0.001$ & $0.033$ & $0.038$ & $0.013$ & $0.007$ & $0.016$ & $0.023$ & $0.031$ & $0.006$ & $0.006$ \\
  \hline
  \multicolumn{11}{l}{$n = 200$}\\
  Bias & $-0.005$ & $-0.026$ & $\phantom{-}0.021$ & $-0.000$ & $\phantom{-}0.000$ & $-0.022$ & $\phantom{-}0.017$ & $\phantom{-}0.000$ & $\phantom{-}0.002$ & $-0.004$ \\
  MSE  & $0.000$ & $0.014$ & $0.018$ & $0.007$ & $0.004$ & $0.006$ & $0.010$ & $0.016$ & $0.003$ & $0.003$ \\
  \hline
  \multicolumn{11}{l}{$n = 500$}\\
  Bias & $-0.002$ & $-0.016$ & $\phantom{-}0.010$ & $\phantom{-}0.004$ & $-0.001$ & $-0.007$ & $\phantom{-}0.003$ & $-0.000$ & $-0.002$ & $-0.002$ \\
  MSE  & $0.000$ & $0.005$ & $0.006$ & $0.003$ & $0.001$ & $0.002$ & $0.004$ & $0.006$ & $0.001$ & $0.001$ \\
  \hline
\end{tabular}}
\end{table}

\begin{table}[!ht]
\footnotesize\centering
\caption{Empirical bias and MSE for the conditional ML estimators of the
  MBSARMA(1,1) model with $\rho = 0.75$ and
  $N_r = 1000$ Monte Carlo replications.}
\label{tab:mc_rho75}
\adjustbox{max width=\linewidth}{\begin{tabular}{l rrrrrrrrrr }
  \hline
  & $\alpha$ & $\phi_{11}$ & $\theta_{11}$ & $\eta_1$ & $\beta_1$ & $\phi_{12}$ & $\theta_{12}$ & $\eta_2$ & $\beta_2$ & $\rho$ \\
  \hline
  True & $0.50$ & $0.50$ & $0.10$ & $1.20$ & $0.30$ & $0.70$ & $0.10$ & $1.20$ & $0.30$ & $0.75$ \\
  \hline
  \multicolumn{11}{l}{$n = 50$}\\
  Bias & $-0.021$ & $-0.099$ & $\phantom{-}0.092$ & $\phantom{-}0.001$ & $\phantom{-}0.001$ & $-0.094$ & $\phantom{-}0.073$ & $-0.002$ & $-0.001$ & $-0.025$ \\
  MSE  & $0.004$ & $0.075$ & $0.090$ & $0.026$ & $0.015$ & $0.040$ & $0.053$ & $0.062$ & $0.012$ & $0.007$ \\
  \hline
  \multicolumn{11}{l}{$n = 100$}\\
  Bias & $-0.010$ & $-0.051$ & $\phantom{-}0.038$ & $\phantom{-}0.001$ & $-0.001$ & $-0.044$ & $\phantom{-}0.032$ & $\phantom{-}0.006$ & $-0.001$ & $-0.010$ \\
  MSE  & $0.001$ & $0.033$ & $0.041$ & $0.013$ & $0.007$ & $0.014$ & $0.022$ & $0.031$ & $0.007$ & $0.002$ \\
  \hline
  \multicolumn{11}{l}{$n = 200$}\\
  Bias & $-0.005$ & $-0.021$ & $\phantom{-}0.018$ & $-0.004$ & $\phantom{-}0.002$ & $-0.019$ & $\phantom{-}0.011$ & $-0.008$ & $-0.000$ & $-0.004$ \\
  MSE  & $0.001$ & $0.014$ & $0.018$ & $0.007$ & $0.003$ & $0.006$ & $0.010$ & $0.016$ & $0.003$ & $0.001$ \\
  \hline
  \multicolumn{11}{l}{$n = 500$}\\
  Bias & $-0.002$ & $-0.005$ & $\phantom{-}0.004$ & $-0.001$ & $-0.002$ & $-0.009$ & $\phantom{-}0.008$ & $-0.001$ & $-0.001$ & $-0.001$ \\
  MSE  & $0.000$ & $0.004$ & $0.006$ & $0.003$ & $0.001$ & $0.002$ & $0.003$ & $0.007$ & $0.001$ & $0.000$ \\
  \hline
\end{tabular}}
\end{table}

\begin{table}[!ht]
\footnotesize\centering
\caption{Empirical bias and MSE for the conditional ML estimators of the
  MBSARMA(1,1) model with $\rho = 0.90$ and
  $N_r = 1000$ Monte Carlo replications.}
\label{tab:mc_rho90}
\adjustbox{max width=\linewidth}{\begin{tabular}{l rrrrrrrrrr }
  \hline
  & $\alpha$ & $\phi_{11}$ & $\theta_{11}$ & $\eta_1$ & $\beta_1$ & $\phi_{12}$ & $\theta_{12}$ & $\eta_2$ & $\beta_2$ & $\rho$ \\
  \hline
  True & $0.50$ & $0.50$ & $0.10$ & $1.20$ & $0.30$ & $0.70$ & $0.10$ & $1.20$ & $0.30$ & $0.90$ \\
  \hline
  \multicolumn{11}{l}{$n = 50$}\\
  Bias & $-0.022$ & $-0.107$ & $\phantom{-}0.094$ & $-0.013$ & $-0.000$ & $-0.096$ & $\phantom{-}0.076$ & $-0.025$ & $-0.002$ & $-0.020$ \\
  MSE  & $0.003$ & $0.079$ & $0.095$ & $0.028$ & $0.016$ & $0.043$ & $0.059$ & $0.070$ & $0.013$ & $0.003$ \\
  \hline
  \multicolumn{11}{l}{$n = 100$}\\
  Bias & $-0.011$ & $-0.041$ & $\phantom{-}0.033$ & $-0.002$ & $-0.003$ & $-0.041$ & $\phantom{-}0.031$ & $-0.003$ & $-0.002$ & $-0.006$ \\
  MSE  & $0.001$ & $0.030$ & $0.040$ & $0.013$ & $0.007$ & $0.013$ & $0.022$ & $0.032$ & $0.006$ & $0.001$ \\
  \hline
  \multicolumn{11}{l}{$n = 200$}\\
  Bias & $-0.006$ & $-0.027$ & $\phantom{-}0.022$ & $\phantom{-}0.000$ & $\phantom{-}0.003$ & $-0.022$ & $\phantom{-}0.015$ & $\phantom{-}0.004$ & $\phantom{-}0.003$ & $-0.003$ \\
  MSE  & $0.001$ & $0.014$ & $0.017$ & $0.007$ & $0.003$ & $0.006$ & $0.010$ & $0.016$ & $0.003$ & $0.000$ \\
  \hline
  \multicolumn{11}{l}{$n = 500$}\\
  Bias & $-0.003$ & $-0.007$ & $\phantom{-}0.005$ & $\phantom{-}0.001$ & $-0.000$ & $-0.009$ & $\phantom{-}0.006$ & $\phantom{-}0.002$ & $\phantom{-}0.000$ & $-0.001$ \\
  MSE  & $0.000$ & $0.005$ & $0.006$ & $0.003$ & $0.001$ & $0.002$ & $0.004$ & $0.007$ & $0.001$ & $0.000$ \\
  \hline
\end{tabular}}
\end{table}

\begin{table}[!ht]
\footnotesize\centering
\caption{Empirical bias and MSE for the conditional ML estimators of the
  MBSARMA(1,1) model with $d = 3$ components, $\rho = 0.25$ and
  $N_r = 1000$ Monte Carlo replications.}
\label{tab:mc3_rho25}
{(a) Dynamic parameters}\\[4pt]
\adjustbox{max width=\linewidth}{\begin{tabular}{l rrrrr rrrrr rrrr}
  \hline
  & $\alpha$ & $\phi_{11}$ & $\theta_{11}$ & $\eta_1$ & $\beta_1$
  & $\phi_{12}$ & $\theta_{12}$ & $\eta_2$ & $\beta_2$
  & $\phi_{13}$ & $\theta_{13}$ & $\eta_3$ & $\beta_3$ \\
  \hline
  True & $0.50$ & $0.50$ & $0.10$ & $1.20$ & $0.30$ & $0.70$ & $0.10$ & $1.20$ & $0.30$ & $0.60$ & $0.10$ & $1.20$ & $0.30$ \\
  \hline
  \multicolumn{14}{l}{$n = 50$}\\
  Bias & $-0.021$ & $-0.111$ & $\phantom{-}0.096$ & $-0.003$ & $-0.000$ & $-0.100$ & $\phantom{-}0.081$ & $-0.010$ & $-0.001$ & $-0.118$ & $\phantom{-}0.102$ & $-0.003$ & $\phantom{-}0.006$ \\
  MSE  & $0.003$ & $0.088$ & $0.098$ & $0.026$ & $0.016$ & $0.044$ & $0.065$ & $0.060$ & $0.014$ & $0.069$ & $0.084$ & $0.036$ & $0.015$ \\
  \hline
  \multicolumn{14}{l}{$n = 100$}\\
  Bias & $-0.012$ & $-0.053$ & $\phantom{-}0.041$ & $\phantom{-}0.007$ & $-0.004$ & $-0.047$ & $\phantom{-}0.040$ & $-0.001$ & $-0.003$ & $-0.051$ & $\phantom{-}0.035$ & $\phantom{-}0.007$ & $-0.002$ \\
  MSE  & $0.001$ & $0.034$ & $0.037$ & $0.014$ & $0.007$ & $0.014$ & $0.023$ & $0.031$ & $0.006$ & $0.022$ & $0.028$ & $0.019$ & $0.006$ \\
  \hline
  \multicolumn{14}{l}{$n = 200$}\\
  Bias & $-0.005$ & $-0.024$ & $\phantom{-}0.018$ & $-0.000$ & $\phantom{-}0.003$ & $-0.025$ & $\phantom{-}0.017$ & $\phantom{-}0.003$ & $\phantom{-}0.002$ & $-0.028$ & $\phantom{-}0.020$ & $\phantom{-}0.002$ & $\phantom{-}0.001$ \\
  MSE  & $0.000$ & $0.013$ & $0.017$ & $0.006$ & $0.004$ & $0.006$ & $0.010$ & $0.017$ & $0.003$ & $0.010$ & $0.013$ & $0.010$ & $0.003$ \\
  \hline
  \multicolumn{14}{l}{$n = 500$}\\
  Bias & $-0.002$ & $-0.008$ & $\phantom{-}0.005$ & $\phantom{-}0.001$ & $-0.001$ & $-0.009$ & $\phantom{-}0.006$ & $-0.005$ & $\phantom{-}0.000$ & $-0.010$ & $\phantom{-}0.007$ & $-0.003$ & $\phantom{-}0.000$ \\
  MSE  & $0.000$ & $0.005$ & $0.006$ & $0.003$ & $0.001$ & $0.002$ & $0.004$ & $0.006$ & $0.001$ & $0.003$ & $0.005$ & $0.004$ & $0.001$ \\
  \hline
\end{tabular}}

\medskip
{(b) Correlation parameters}\\[4pt]
\adjustbox{max width=\linewidth}{\begin{tabular}{l rrr}
  \hline
  & $\rho_{12}$ & $\rho_{13}$ & $\rho_{23}$ \\
  \hline
  True & $0.25$ & $0.25$ & $0.25$ \\
  \hline
  \multicolumn{4}{l}{$n = 50$}\\
  Bias & $-0.017$ & $-0.022$ & $-0.020$ \\
  MSE  & $0.019$ & $0.020$ & $0.021$ \\
  \hline
  \multicolumn{4}{l}{$n = 100$}\\
  Bias & $-0.009$ & $-0.010$ & $-0.006$ \\
  MSE  & $0.010$ & $0.010$ & $0.009$ \\
  \hline
  \multicolumn{4}{l}{$n = 200$}\\
  Bias & $-0.005$ & $-0.003$ & $-0.003$ \\
  MSE  & $0.005$ & $0.004$ & $0.004$ \\
  \hline
  \multicolumn{4}{l}{$n = 500$}\\
  Bias & $\phantom{-}0.000$ & $\phantom{-}0.001$ & $-0.001$ \\
  MSE  & $0.002$ & $0.002$ & $0.002$ \\
  \hline
\end{tabular}}
\end{table}

\begin{table}[!ht]
\footnotesize\centering
\caption{Empirical bias and MSE for the conditional ML estimators of the
  MBSARMA(1,1) model with $d = 3$ components, $\rho = 0.50$ and
  $N_r = 1000$ Monte Carlo replications.}
\label{tab:mc3_rho50}
{(a) Dynamic parameters}\\[4pt]
\adjustbox{max width=\linewidth}{\begin{tabular}{l rrrrr rrrrr rrrr}
  \hline
  & $\alpha$ & $\phi_{11}$ & $\theta_{11}$ & $\eta_1$ & $\beta_1$
  & $\phi_{12}$ & $\theta_{12}$ & $\eta_2$ & $\beta_2$
  & $\phi_{13}$ & $\theta_{13}$ & $\eta_3$ & $\beta_3$ \\
  \hline
  True & $0.50$ & $0.50$ & $0.10$ & $1.20$ & $0.30$ & $0.70$ & $0.10$ & $1.20$ & $0.30$ & $0.60$ & $0.10$ & $1.20$ & $0.30$ \\
  \hline
  \multicolumn{14}{l}{$n = 50$}\\
  Bias & $-0.022$ & $-0.113$ & $\phantom{-}0.101$ & $\phantom{-}0.009$ & $-0.008$ & $-0.097$ & $\phantom{-}0.074$ & $-0.007$ & $-0.002$ & $-0.100$ & $\phantom{-}0.088$ & $-0.005$ & $\phantom{-}0.000$ \\
  MSE  & $0.002$ & $0.083$ & $0.096$ & $0.024$ & $0.015$ & $0.042$ & $0.057$ & $0.063$ & $0.012$ & $0.060$ & $0.081$ & $0.039$ & $0.014$ \\
  \hline
  \multicolumn{14}{l}{$n = 100$}\\
  Bias & $-0.012$ & $-0.058$ & $\phantom{-}0.049$ & $-0.001$ & $\phantom{-}0.002$ & $-0.049$ & $\phantom{-}0.040$ & $\phantom{-}0.008$ & $\phantom{-}0.005$ & $-0.041$ & $\phantom{-}0.037$ & $-0.001$ & $\phantom{-}0.001$ \\
  MSE  & $0.001$ & $0.032$ & $0.038$ & $0.014$ & $0.007$ & $0.014$ & $0.022$ & $0.033$ & $0.006$ & $0.019$ & $0.028$ & $0.019$ & $0.007$ \\
  \hline
  \multicolumn{14}{l}{$n = 200$}\\
  Bias & $-0.006$ & $-0.032$ & $\phantom{-}0.030$ & $-0.007$ & $\phantom{-}0.001$ & $-0.026$ & $\phantom{-}0.019$ & $-0.006$ & $\phantom{-}0.000$ & $-0.025$ & $\phantom{-}0.020$ & $-0.002$ & $\phantom{-}0.000$ \\
  MSE  & $0.000$ & $0.015$ & $0.018$ & $0.007$ & $0.003$ & $0.006$ & $0.010$ & $0.016$ & $0.003$ & $0.009$ & $0.013$ & $0.009$ & $0.003$ \\
  \hline
  \multicolumn{14}{l}{$n = 500$}\\
  Bias & $-0.002$ & $-0.009$ & $\phantom{-}0.006$ & $\phantom{-}0.002$ & $\phantom{-}0.000$ & $-0.009$ & $\phantom{-}0.008$ & $\phantom{-}0.001$ & $\phantom{-}0.001$ & $-0.007$ & $\phantom{-}0.007$ & $-0.001$ & $\phantom{-}0.001$ \\
  MSE  & $0.000$ & $0.005$ & $0.006$ & $0.003$ & $0.001$ & $0.002$ & $0.004$ & $0.006$ & $0.001$ & $0.003$ & $0.005$ & $0.004$ & $0.001$ \\
  \hline
\end{tabular}}

\medskip
{(b) Correlation parameters}\\[4pt]
\adjustbox{max width=\linewidth}{\begin{tabular}{l rrr}
  \hline
  & $\rho_{12}$ & $\rho_{13}$ & $\rho_{23}$ \\
  \hline
  True & $0.50$ & $0.50$ & $0.50$ \\
  \hline
  \multicolumn{4}{l}{$n = 50$}\\
  Bias & $-0.025$ & $-0.029$ & $-0.028$ \\
  MSE  & $0.014$ & $0.016$ & $0.015$ \\
  \hline
  \multicolumn{4}{l}{$n = 100$}\\
  Bias & $-0.008$ & $-0.011$ & $-0.012$ \\
  MSE  & $0.006$ & $0.006$ & $0.006$ \\
  \hline
  \multicolumn{4}{l}{$n = 200$}\\
  Bias & $-0.008$ & $-0.008$ & $-0.007$ \\
  MSE  & $0.003$ & $0.003$ & $0.003$ \\
  \hline
  \multicolumn{4}{l}{$n = 500$}\\
  Bias & $-0.003$ & $-0.002$ & $-0.002$ \\
  MSE  & $0.001$ & $0.001$ & $0.001$ \\
  \hline
\end{tabular}}
\end{table}

\begin{table}[!ht]
\footnotesize\centering
\caption{Empirical bias and MSE for the conditional ML estimators of the
  MBSARMA(1,1) model with $d = 3$ components, $\rho = 0.75$ and
  $N_r = 1000$ Monte Carlo replications.}
\label{tab:mc3_rho75}
{(a) Dynamic parameters}\\[4pt]
\adjustbox{max width=\linewidth}{\begin{tabular}{l rrrrr rrrrr rrrr}
  \hline
  & $\alpha$ & $\phi_{11}$ & $\theta_{11}$ & $\eta_1$ & $\beta_1$
  & $\phi_{12}$ & $\theta_{12}$ & $\eta_2$ & $\beta_2$
  & $\phi_{13}$ & $\theta_{13}$ & $\eta_3$ & $\beta_3$ \\
  \hline
  True & $0.50$ & $0.50$ & $0.10$ & $1.20$ & $0.30$ & $0.70$ & $0.10$ & $1.20$ & $0.30$ & $0.60$ & $0.10$ & $1.20$ & $0.30$ \\
  \hline
  \multicolumn{14}{l}{$n = 50$}\\
  Bias & $-0.020$ & $-0.105$ & $\phantom{-}0.090$ & $\phantom{-}0.010$ & $-0.000$ & $-0.090$ & $\phantom{-}0.077$ & $\phantom{-}0.003$ & $-0.001$ & $-0.099$ & $\phantom{-}0.084$ & $\phantom{-}0.002$ & $\phantom{-}0.003$ \\
  MSE  & $0.002$ & $0.078$ & $0.096$ & $0.025$ & $0.015$ & $0.040$ & $0.062$ & $0.062$ & $0.012$ & $0.058$ & $0.073$ & $0.039$ & $0.014$ \\
  \hline
  \multicolumn{14}{l}{$n = 100$}\\
  Bias & $-0.011$ & $-0.047$ & $\phantom{-}0.040$ & $-0.003$ & $\phantom{-}0.002$ & $-0.040$ & $\phantom{-}0.035$ & $-0.000$ & $\phantom{-}0.001$ & $-0.047$ & $\phantom{-}0.044$ & $\phantom{-}0.002$ & $\phantom{-}0.001$ \\
  MSE  & $0.001$ & $0.035$ & $0.042$ & $0.013$ & $0.007$ & $0.014$ & $0.021$ & $0.034$ & $0.006$ & $0.022$ & $0.030$ & $0.019$ & $0.006$ \\
  \hline
  \multicolumn{14}{l}{$n = 200$}\\
  Bias & $-0.006$ & $-0.023$ & $\phantom{-}0.018$ & $-0.004$ & $-0.002$ & $-0.023$ & $\phantom{-}0.018$ & $-0.009$ & $\phantom{-}0.001$ & $-0.021$ & $\phantom{-}0.015$ & $-0.005$ & $\phantom{-}0.002$ \\
  MSE  & $0.000$ & $0.013$ & $0.016$ & $0.006$ & $0.003$ & $0.006$ & $0.010$ & $0.016$ & $0.003$ & $0.008$ & $0.012$ & $0.010$ & $0.003$ \\
  \hline
  \multicolumn{14}{l}{$n = 500$}\\
  Bias & $-0.002$ & $-0.007$ & $\phantom{-}0.004$ & $-0.000$ & $\phantom{-}0.002$ & $-0.007$ & $\phantom{-}0.004$ & $-0.000$ & $\phantom{-}0.001$ & $-0.007$ & $\phantom{-}0.004$ & $\phantom{-}0.001$ & $\phantom{-}0.002$ \\
  MSE  & $0.000$ & $0.005$ & $0.006$ & $0.003$ & $0.001$ & $0.002$ & $0.004$ & $0.007$ & $0.001$ & $0.003$ & $0.005$ & $0.004$ & $0.001$ \\
  \hline
\end{tabular}}

\medskip
{(b) Correlation parameters}\\[4pt]
\adjustbox{max width=\linewidth}{\begin{tabular}{l rrr}
  \hline
  & $\rho_{12}$ & $\rho_{13}$ & $\rho_{23}$ \\
  \hline
  True & $0.75$ & $0.75$ & $0.75$ \\
  \hline
  \multicolumn{4}{l}{$n = 50$}\\
  Bias & $-0.027$ & $-0.023$ & $-0.021$ \\
  MSE  & $0.007$ & $0.006$ & $0.006$ \\
  \hline
  \multicolumn{4}{l}{$n = 100$}\\
  Bias & $-0.011$ & $-0.009$ & $-0.009$ \\
  MSE  & $0.002$ & $0.002$ & $0.002$ \\
  \hline
  \multicolumn{4}{l}{$n = 200$}\\
  Bias & $-0.006$ & $-0.004$ & $-0.004$ \\
  MSE  & $0.001$ & $0.001$ & $0.001$ \\
  \hline
  \multicolumn{4}{l}{$n = 500$}\\
  Bias & $-0.002$ & $-0.001$ & $-0.002$ \\
  MSE  & $0.000$ & $0.000$ & $0.000$ \\
  \hline
\end{tabular}}
\end{table}

\setlength{\floatsep}{12pt plus 2pt minus 2pt}%
\setlength{\textfloatsep}{20pt plus 2pt minus 4pt}%

\clearpage

\section{Application to air pollution data}\label{sec:05}

In order to illustrate the practical utility of the proposed MBSARMA model,
we analyze weekly PM$_{2.5}$ concentration data recorded at three monitoring
stations in Santiago, Chile. Fine particulate matter concentrations are
defined on the positive real line and exhibit positive skewness and
heavy-tailed behavior, features that are consistent with the
Birnbaum--Saunders marginal distribution underlying the proposed model. This
distributional characterization has been documented in the recent literature:
\cite{puentes:21} showed that bivariate BS log-linear models effectively
predict PM$_{2.5}$ and PM$_{10}$ levels during critical pollution episodes in
Santiago, Chile, and \cite{ibacache:26} confirmed that PM$_{2.5}$
concentrations exhibit characteristics of the BS distribution, namely positive
skewness and non-negative support. The data also exhibit strong
contemporaneous spatial dependence across stations, further motivating the use
of the multivariate BS distributional framework.


We analyze weekly mean PM$_{2.5}$ concentrations
recorded at three monitoring stations of the Red MACAM network in the
Metropolitan Region of Santiago, Chile. The data were obtained from the
National Air Quality Information System of Chile
\citep[SINCA;][]{sinca:mma}, which provides hourly pollutant measurements at
stations across the country. We consider three stations that cover distinct
geographical sectors of the city: Las Condes (eastern sector), El Bosque
(southern sector), and Quilicura (northern sector). Santiago is a major
South American metropolitan area where PM$_{2.5}$ concentrations are strongly
shaped by wintertime temperature inversions \citep{munoz:23}, making the
joint modeling of multiple stations an important task. The study
period spans April 2016 to December 2017, yielding $n = 82$ weekly
observations after retaining only weeks with at least $90$ valid hourly
records across all three stations.

The series are denoted $S_{t1}$, $S_{t2}$, and $S_{t3}$, representing weekly
mean PM$_{2.5}$ at Las Condes, El Bosque, and Quilicura, respectively.
The log-transformed responses $Y_{tj} = \log(S_{tj})$ are considered
throughout. Five exogenous covariates are included ($k = 5$): a normalised linear trend
$x_{t1} = t/n$ and two pairs of
harmonic terms capturing annual and semi-annual pollution cycles driven by
wintertime temperature inversions and meteorological conditions.

\begin{figure}[ht]
\centering
\includegraphics[width=\textwidth]{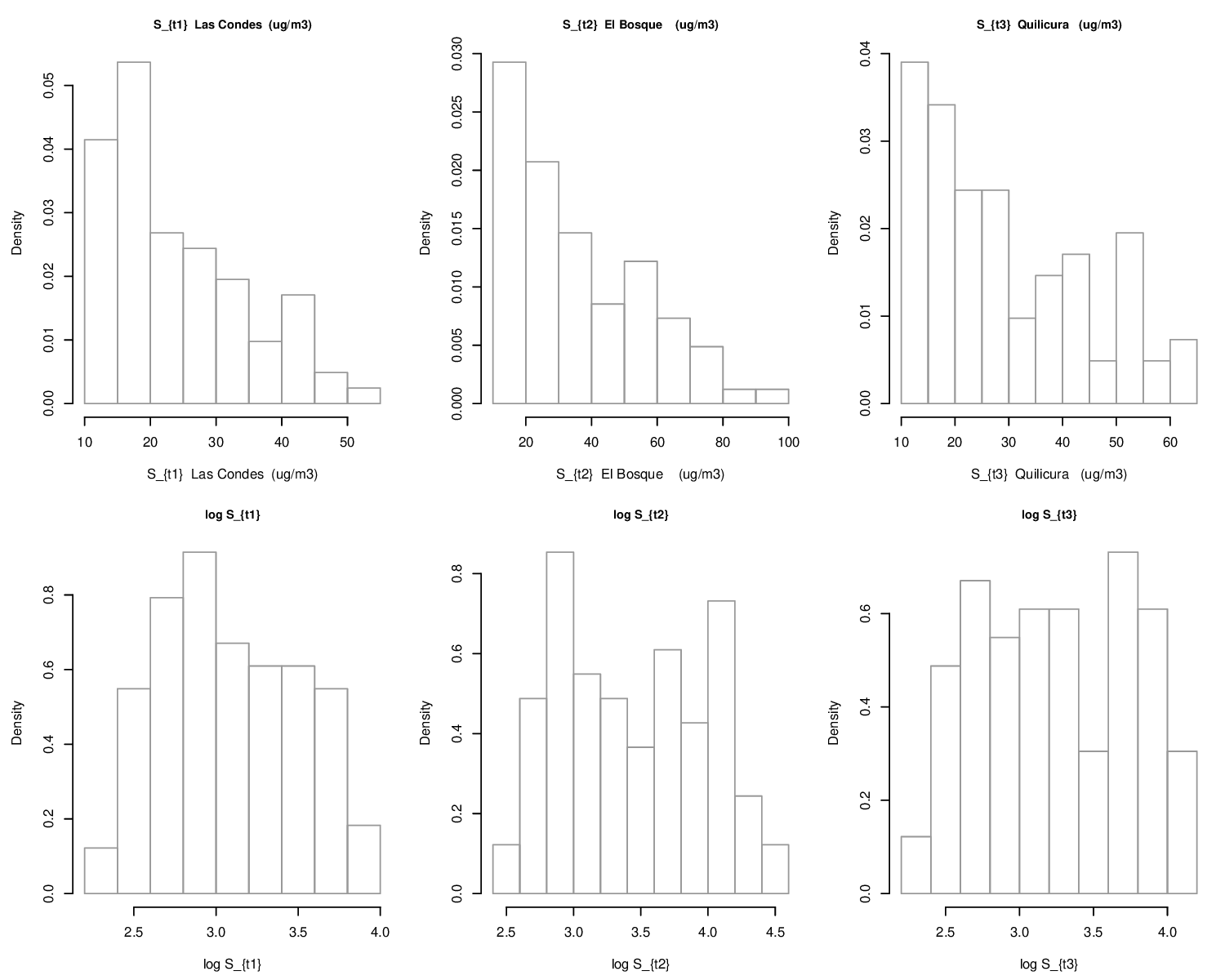}
\caption{Histograms of the weekly PM$_{2.5}$ series (top row, raw scale)
  and their log-transforms (bottom row) for Las Condes ($S_{t1}$, left),
  El Bosque ($S_{t2}$, centre), and Quilicura ($S_{t3}$, right).}
\label{fig:hist2}
\end{figure}

Table~\ref{tab:descr2} reports summary statistics for all three series on
the original and log scale. All three raw series exhibit positive skewness
(ranging from $0.602$ to $0.831$) and kurtosis between $2$ and $3$ (ranging from $2.085$ to
$2.853$), consistent with the Birnbaum--Saunders marginal distribution
underlying the MBSARMA model. The log-transformed series are approximately
symmetric, with skewness close to zero and kurtosis near $2$.
The pairwise Pearson correlations between the log-transformed series are
$0.847$ ($S_{t1}$--$S_{t2}$), $0.903$ ($S_{t1}$--$S_{t3}$), and $0.957$
($S_{t2}$--$S_{t3}$). Figure~\ref{fig:hist2} shows histograms of all six series, confirming the pronounced positive
skewness of the raw concentrations.

\begin{table}[ht]
\small\centering
\caption{Descriptive statistics for the weekly PM$_{2.5}$ series
  (Las Condes, El Bosque, and Quilicura stations, Santiago).}
\label{tab:descr2}
\begin{tabular}{lccccccccc}
  \hline
  Series & $n$ & Mean & Median & SD & CV & Skewness & Kurtosis & Min & Max \\
  \hline
  $S_{t1}$ (Las Condes)  & $82$ & $23.98$ & $21.15$ & $10.20$ & $0.425$ & $0.734$ & $2.490$ & $10.77$ & $51.74$ \\
  $S_{t2}$ (El Bosque)   & $82$ & $36.19$ & $30.54$ & $19.81$ & $0.547$ & $0.831$ & $2.853$ & $12.40$ & $94.22$ \\
  $S_{t3}$ (Quilicura)   & $82$ & $29.31$ & $25.93$ & $15.06$ & $0.514$ & $0.602$ & $2.085$ & $10.18$ & $64.05$ \\
  $\log S_{t1}$          & $82$ & $3.09$ & $3.05$ & $0.41$ & $0.134$ & $0.177$ & $1.913$ & $2.38$ & $3.95$ \\
  $\log S_{t2}$          & $82$ & $3.44$ & $3.42$ & $0.54$ & $0.158$ & $0.119$ & $1.784$ & $2.52$ & $4.55$ \\
  $\log S_{t3}$          & $82$ & $3.25$ & $3.25$ & $0.52$ & $0.161$ & $0.044$ & $1.726$ & $2.32$ & $4.16$ \\
  \hline
\end{tabular}
\end{table}

All three log-transformed series exhibit moderate autocorrelation at lag~1
($\widehat{\rho}_1 \approx 0.500$ for $\log S_{t1}$, $0.787$ for $\log S_{t2}$,
and $0.690$ for $\log S_{t3}$), and the Ljung--Box test at lag~12 rejects
the white-noise hypothesis for all three series ($p < 0.05$), motivating
the inclusion of ARMA components.

We fit MBSARMA($\mathbf{p},\mathbf{q}$) models to the data for several
choices of AR and MA orders and select the best-fitting model based on
the Bayesian information criterion
(BIC) 
\citep{schwarz:78}, given by
\begin{equation}\label{eq:ic}
  \text{BIC} = -2\ell(\widehat{\boldsymbol{\gamma}}) + \kappa\log(n-m),
\end{equation}
where $\kappa$ is the total number of free parameters and
$\ell(\widehat{\boldsymbol{\gamma}})$ is the maximized conditional log-likelihood.
We consider symmetric MBSARMA orders $(p,p,p)$ and $(q,q,q)$ for $p,q \in
\{0,1,2\}$ as well as selected asymmetric combinations, evaluated on
$n_{\text{train}} = n - 8$ training observations (the last $8$ weeks are
reserved as a test set).
The best BIC ($-63.30$) was achieved by the MBSARMA$(0,0,0|1,1,1)$
model among all candidates considered.

Table~\ref{tab:app2} reports the conditional ML estimates and corresponding standard
errors for the selected model. The estimated shape parameter $\widehat{\alpha} = 0.299$
reflects moderate variability in the weekly PM$_{2.5}$ concentrations over the
study period. The MA coefficients $\widehat{\theta}_{11} = 0.205$, $\widehat{\theta}_{12} = 0.207$,
and $\widehat{\theta}_{13} = 0.126$ are positive and moderate, indicating short-memory
dynamics in the conditional location. The negative trend
coefficients $\widehat{\beta}_{1j} < 0$ are consistent with a gradual reduction in
PM$_{2.5}$ levels over the study period. The pairwise correlation
estimates $\widehat{\rho}_{12} = 0.864$, $\widehat{\rho}_{13} = 0.918$, and
$\widehat{\rho}_{23} = 0.913$ confirm strong correlation across the three
monitoring stations, with the highest correlation between Las Condes (east) and
Quilicura (north), both of which are more heavily affected by wintertime thermal
inversions \citep{munoz:23}.

\begin{table}[!ht]
\small\centering
\caption{Conditional ML estimates and standard errors (SE) for the
  MBSARMA$(0,0,0|1,1,1)$ model fitted to the weekly PM$_{2.5}$ series
  (Las Condes, El Bosque, Quilicura, Santiago;
  $n_{\text{train}} = 74$ weeks,
  $\ell(\widehat{\boldsymbol{\gamma}}) = 85.28$,
  $\text{BIC} = -63.30$).
  Covariates: $\widehat{\beta}_{1j}$ = trend,
  $\widehat{\beta}_{2j}$ = $\sin(2\pi w/52)$, $\widehat{\beta}_{3j}$ = $\cos(2\pi w/52)$,
  $\widehat{\beta}_{4j}$ = $\sin(4\pi w/52)$, $\widehat{\beta}_{5j}$ = $\cos(4\pi w/52)$.}
\label{tab:app2}
\begin{tabular}{lrrrrrrr}
  \hline
  & $\widehat{\theta}_{1j}$ & $\widehat{\eta}_j$
  & $\widehat{\beta}_{1j}$ & $\widehat{\beta}_{2j}$ & $\widehat{\beta}_{3j}$
  & $\widehat{\beta}_{4j}$ & $\widehat{\beta}_{5j}$ \\
  \hline
  $j=1$ (Las Condes)
    & $0.2046$ & $3.2684$ & $-0.4079$ & $0.0105$ & $-0.2050$ & $0.0937$ & $0.1705$ \\
  SE
    & $(0.0675)$ & $(0.0901)$ & $(0.1675)$ & $(0.0647)$ & $(0.0592)$ & $(0.0584)$ & $(0.0635)$ \\
  \hline
  $j=2$ (El Bosque)
    & $0.2070$ & $3.4866$ & $-0.2012$ & $0.0730$ & $-0.5101$ & $0.0754$ & $0.2869$ \\
  SE
    & $(0.0718)$ & $(0.0912)$ & $(0.1692)$ & $(0.0653)$ & $(0.0600)$ & $(0.0591)$ & $(0.0641)$ \\
  \hline
  $j=3$ (Quilicura)
    & $0.1261$ & $3.3201$ & $-0.2168$ & $0.0928$ & $-0.3682$ & $0.0731$ & $0.3458$ \\
  SE
    & $(0.0634)$ & $(0.0844)$ & $(0.1570)$ & $(0.0606)$ & $(0.0556)$ & $(0.0550)$ & $(0.0595)$ \\
  \hline
\end{tabular}
\medskip

\begin{tabular}{lrrrr}
  \hline
  & $\widehat{\alpha}$ & $\widehat{\rho}_{12}$ & $\widehat{\rho}_{13}$ & $\widehat{\rho}_{23}$ \\
  \hline
  Estimate & $0.2986$ & $0.8642$ & $0.9183$ & $0.9134$ \\
  SE       & $(0.0216)$ & $(0.0264)$ & $(0.0159)$ & $(0.0192)$ \\
  \hline
\end{tabular}
\end{table}

\begin{figure}[!ht]
\centering
\includegraphics[width=0.55\textwidth]{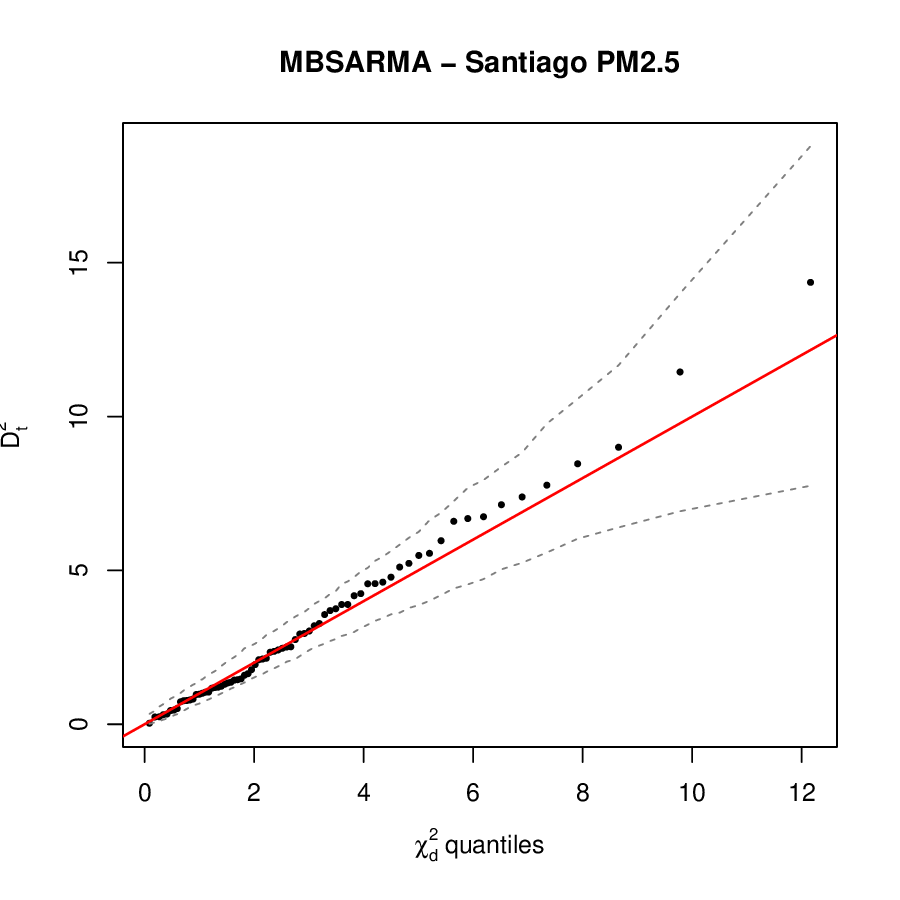}
\caption{Chi-squared ($\chi^2_3$) QQ plot of the Mahalanobis distance
  residuals $D_t^2$ with $95\%$ simulated envelopes.}
\label{fig:qqresid2}
\end{figure}

In order to assess the goodness of fit of the selected model, we compute the
Mahalanobis distance residuals $D_t^2$ defined in~(\ref{eq:mahal}), which
under correct specification follow a $\chi^2_d$ distribution with $d = 3$.
Figure~\ref{fig:qqresid2} shows the $\chi^2_3$ QQ plot of the $D_t^2$ values
against simulated envelopes. We see clearly that the MBSARMA model provides a good fit for these data. A Kolmogorov--Smirnov test yields
$D = 0.072$ (p-value $= 0.821$), providing no evidence against the $\chi^2_3$
null distribution at the $5\%$ level and confirming the good fit of the selected model.

\begin{figure}[!ht]
\centering
\includegraphics[width=\textwidth]{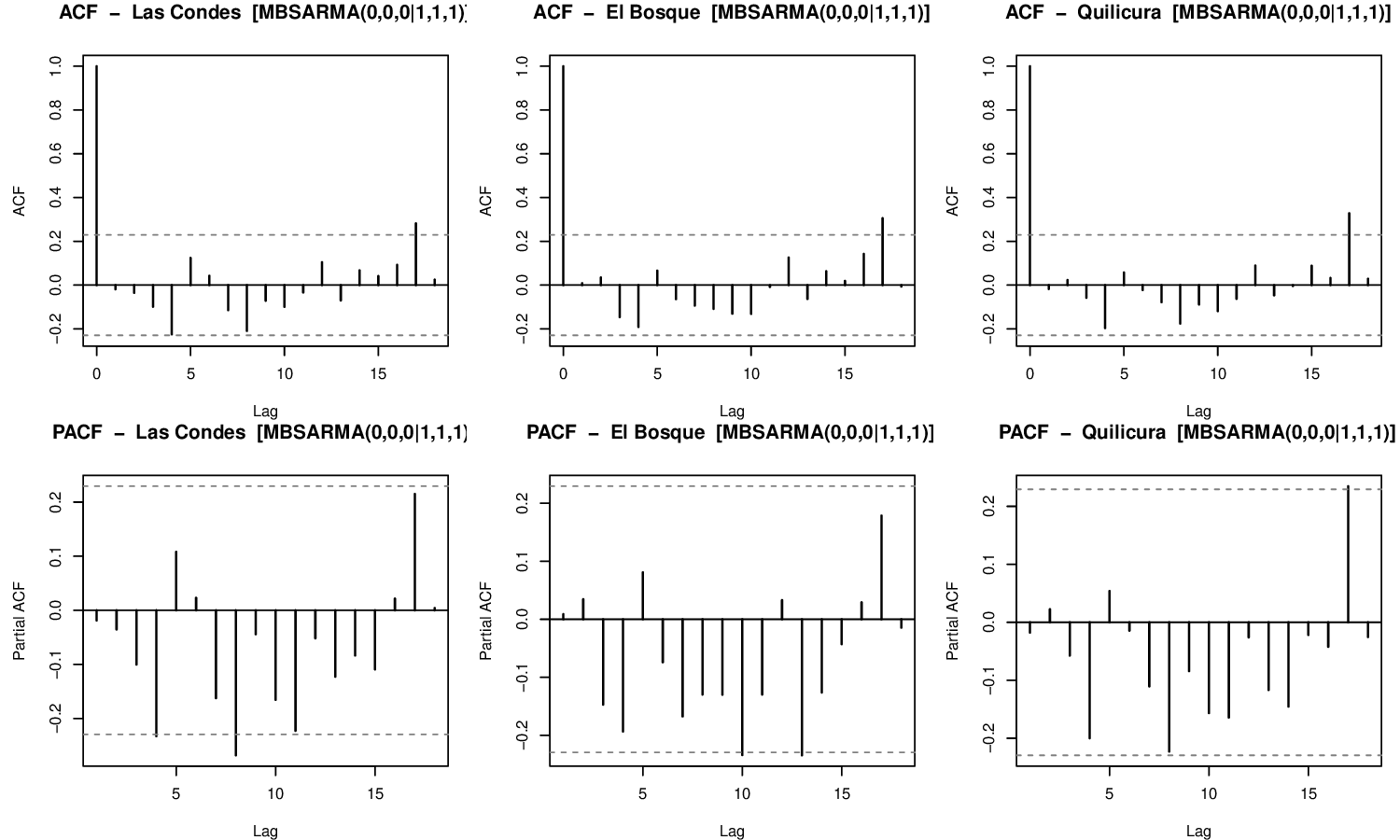}
\caption{Sample ACF (top row) and PACF (bottom row) of the component residuals
  $\widehat{a}_{tj}$ under the fitted MBSARMA$(0,0,0|1,1,1)$ model for
  Las Condes ($j=1$, left), El Bosque ($j=2$, centre), and
  Quilicura ($j=3$, right).}
\label{fig:residacf2}
\end{figure}

To further assess the adequacy of the fitted ARMA structure, Figure~\ref{fig:residacf2}
displays the sample ACF and PACF of the component residuals
$\widehat{a}_{tj} = (2/\widehat{\alpha})\sinh\bigl((\widehat{Y}_{tj} - \widehat{\mu}_{tj})/2\bigr)$
for each of the three stations. In general, the plots indicate that the selected MBSARMA model
has successfully captured the temporal dependence structure in the data.

To evaluate the forecasting ability of the proposed model, we reserve
the last $8$ weeks of the series as a test set ($n_{\text{test}} = 8$) and
compute one-step-ahead predictions using~(\ref{eq:pred1})
and~(\ref{eq:tau_pred}). Table~\ref{tab:rmse2} compares the RMSE and MAE of
the selected MBSARMA$(0,0,0|1,1,1)$ model against a Gaussian ARMAX$(0,1)$
benchmark fitted separately to each component using the same set of
covariates, and against the na\"{i}ve predictor, which repeats the last
observed log-concentration for all forecast horizons, that is,
$\widehat{Y}_{n+h,j}^{\text{naïve}} = Y_{n,j}$ for $h = 1, \ldots, 8$. The MBSARMA$(0,0,0|1,1,1)$ model achieves the lowest RMSE and MAE for
$j = 1$ and $j = 2$, and the lowest MAE for $j = 3$, outperforming the
ARMAX$(0,1)$ benchmark across all stations and metrics. For $j = 3$
(Quilicura), the na\"{i}ve predictor attains a slightly lower RMSE than
MBSARMA, whereas the MBSARMA model still yields a lower MAE at that
station. Overall, these results confirm the practical utility of the
proposed MBSARMA model for jointly modelling and forecasting correlated
PM$_{2.5}$ series with positively skewed marginals.

\begin{table}[!ht]
\small\centering
\caption{Root mean squared error (RMSE) and mean absolute error (MAE) for
  one-step-ahead predictions on the $8$-week test set
  (Las Condes, El Bosque, and Quilicura weekly PM$_{2.5}$).
  Na\"{i}ve: $\widehat{Y}_{n+h,j} = Y_{n,j}$ for all $h$.}
\label{tab:rmse2}
\begin{tabular}{lrrrrrr}
  \hline
  & \multicolumn{3}{c}{RMSE} & \multicolumn{3}{c}{MAE} \\
  \cline{2-4}\cline{5-7}
  Model & $Y_{t1}$ & $Y_{t2}$ & $Y_{t3}$ & $Y_{t1}$ & $Y_{t2}$ & $Y_{t3}$ \\
  \hline
  MBSARMA$(0,0,0|1,1,1)$ & $0.2027$ & $0.2179$ & $0.3129$ & $0.1666$ & $0.1777$ & $0.2054$ \\
  ARMAX$(0,1)$ & $1.6427$ & $1.8213$ & $1.8836$ & $1.5943$ & $1.7625$ & $1.8210$ \\
  Na\"ive & $0.2810$ & $0.2867$ & $0.2543$ & $0.2522$ & $0.2350$ & $0.2211$ \\
  \hline
\end{tabular}
\end{table}

\section{Concluding remarks}\label{sec:06}

Multivariate positive time series arise naturally in a wide range of applied
contexts, including financial markets, reliability engineering, and
environmental monitoring, where data are correlated across components,
temporally dependent, and often positively skewed. In this paper, we proposed
the MBSARMA model, which extends the multivariate log-BS framework of
\cite{marchant:15} by incorporating the ARMA dynamics of \cite{leiva:21},
enabling the joint modelling of correlated positive asymmetric time series.
We estimated the model parameters by means of the EM algorithm and Monte Carlo simulation studies were
carried out for evaluating the performance of the estimatorsfor both a bivariate ($d = 2$) and a trivariate ($d = 3$)
setting. Both studies demonstrated that the
conditional ML estimators exhibit good performance, with bias and
MSE decreasing consistently as the sample size increases, across the full
range of correlation structures and dimensions considered. An application to
weekly PM$_{2.5}$ concentrations from three stations of the Red MACAM network
in Santiago, Chile (Las Condes, El Bosque, and Quilicura), obtained from
SINCA \citep{sinca:mma}, confirmed the good fit of the model and
its superior one-step-ahead forecasting performance relative
to univariate competitors, illustrating the applicability of the MBSARMA framework to air
quality monitoring data. As part of future research, it is
of interest to propose heavy-tailed variants based on the BS-$t$ kernel and seasonal MBSARMA
formulations. Work on these problems is currently in progress and we hope to report these findings in future.

\paragraph{Acknowledgments}
The research was supported in part by CNPq and CAPES grants from the Brazilian government.

\normalsize


\end{document}